\documentclass[reprint,aps,prd,nofootinbib,longbibliography,preprintnumbers]{revtex4-1}
\usepackage{amssymb}
\usepackage{setspace}
\usepackage{amsmath}
\usepackage{amsfonts}
\usepackage{hyperref}
\usepackage{graphicx}
\usepackage{float}
\usepackage{subfigure}
\usepackage{color}
\usepackage{wasysym}

\usepackage{feynmp}
\newcommand{\beq} {\begin{equation}}
\newcommand{\eeq} {\end{equation}}
\newcommand{\bea} {\begin{eqnarray}}
\newcommand{\eea} {\end{eqnarray}}

\newcommand{\cd}{c^{\dagger}}
\newcommand{\dg}{\dagger}
\newcommand*\diff{\mathop{}\!\mathrm{d}}
\newcommand{\non}{\nonumber}
\newcommand{\khs}{k_{hs}}
\newcommand{\sgn}{\text{sgn}}

\DeclareMathOperator{\im}{Im}
\DeclareMathOperator{\re}{Re}

\allowdisplaybreaks

\begin{document}
\title{The  Hubbard model on a triangular lattice -- pseudogap due to  spin-density-wave fluctuations}
\author{Mengxing Ye}
\affiliation{School of Physics and Astronomy, University of Minnesota, Minneapolis, MN 55455}
\author{Andrey V. Chubukov}
\affiliation{School of Physics and Astronomy, University of Minnesota, Minneapolis, MN 55455}
\date{\today}
\begin{abstract}
We calculate the fermionic spectral function $A_k (\omega)$ in the spiral spin-density-wave (SDW) state of the Hubbard model on a quasi-2D triangular lattice at small but finite temperature $T$. The spiral SDW order $\Delta (T)$ develops below $T = T_N$ and has momentum ${ \bf K} = (4\pi/3,0)$. We pay special attention to fermions with momenta ${\bf k}$, for which ${\bf k}$ and ${\bf k} + {\bf K}$ are close to Fermi surface in the absence of SDW. At the mean field level,  $A_k (\omega)$  for such fermions has peaks at $\omega = \pm \Delta (T)$ at $T < T_N$ and displays a conventional Fermi liquid behavior at $T > T_N$. We show that this behavior changes qualitatively beyond mean-field due to singular self-energy contributions from thermal fluctuations in a quasi-2D system. We use a non-perturbative eikonal approach and sum up infinite series of thermal self-energy terms. We show that $A_k (\omega)$ shows peak/dip/hump features at $T < T_N$, with the peak position at $\Delta (T)$ and hump position at $\Delta (T=0)$. Above $T_N$, the hump survives up to $T = T_p > T_N$, and in between $T_N$ and $T_p$ the spectral function displays the pseudogap behavior. We show that the difference between $T_p$ and $T_N$  is controlled by the ratio of in-plane and out-of-plane static spin susceptibilities, which determines the combinatoric factors in the diagrammatic series for the self-energy. For certain values of this ratio, $T_p = T_N$, i.e., the pseudogap region collapses. In this last case, thermal fluctuations are logarithmically singular, yet they do not give rise to pseudogap behavior. Our computational method can be used to study pseudogap physics due to thermal fluctuations in other systems.
 \end{abstract}
\maketitle

%========================================================
\section{Introduction}

The pseudogap behavior, observed in several classes of materials, most notably high $T_c$ cuprates, remains one of the mostly debated phenomenon in correlated electron systems.
There are two key scenarios of the pseudogap, each supported by a set of experiments.
One is that the pseudogap is a distinct state of matter with an order parameter, which is either bilinear in fermions (e.g., loop current order~\cite{Varma1997,*Varma1999}), or a four-fermion
composite order  (e.g., a spin nematic~\cite{Kivelson1998,Fradkin2015,Fernandes2012}), or a topological order that cannot be easily expressed via fermionic operators~\cite{Sachdev2018}. Within this scenario, the experimentally detected onset temperature of a pseudogap, $T_p$, is a phase transition temperature.  The other scenario is that the pseudogap is a precursor to an ordered state -- SDW magnetism~\cite{Schmalian1998,*Schmalian1999,Sadovskii1999,Abanov2003,Sedrakyan2010}, superconductivity~\cite{Randeria1998,Millis1998,Berg2007,Chubukov2019}, or both, with the relative strength of the two precursors set by doping (a precursor to SDW is the dominant one at smaller dopings, and a precursor to superconductivity is the dominant one at larger dopings).
Within this scenario,  the system retains a dynamical memory about the underlying order in some temperature range where the order is already destroyed, and this memory gradually fades and disappears at around $T_p$. At around this temperature the behavior of the spectral function crosses-over to that in a (bad) metal. A similar but not equivalent scenario, is for pseudogap as a precursor to Mott physics~\cite{Gull2015}.
The precursor scenario is not strictly orthogonal to the competing order scenario as, e.g., the depletion of the spectral weight at low energies in the antinodal region due to pseudogap formation does enhance the system's tendency to develop a CDW order with axial momenta, consistent with the one observed in the cuprates~\cite{Metlitski2010,WangYuxuan2014,Chowdhury2014,Atkinson2015}. The same holds for pair density-wave order~\cite{WangYuxuan2015prl,*WangYuxuan2015prb,Agterberg2019}. Whether a pre-existing pseudogap helps the system to develop a topological order is less clear.

In this paper we analyze several aspects of the precursor scenario. There is no clear path to get a precursor behavior at $T=0$, but earlier works~\cite{Schmalian1998,*Schmalian1999,Sedrakyan2010,Tchernyshyov1997,Berg2007,Sadovskii1999} have found that thermal (static) SDW and/or superconducting fluctuations do give rise to precursors and associated pseudogap behavior. In particular, previous studies of quasi-2D systems on a square lattice have found that the pseudogap does develop in some $T$ range above the critical $T_N$ towards a commensurate $(\pi,\pi)$ SDW order~\cite{Schmalian1998,*Schmalian1999,Sedrakyan2010}. {\emph{The question we address is whether this is a generic property of a system near an ordered state, or there are situations when thermal fluctuations are logarithmically singular, but do not give rise to pseudogap behavior.}} To analyze this, lowest-order perturbation theory is not sufficient, and one has to sum up infinite series of singular self-energy corrections due to thermal  SDW fluctuations. There is a well established computational procedure for this, similar to eikonal approximation in the scattering theory~\cite{Eikonal}.
Here we consider, within the same computational scheme, the effects of thermal SDW fluctuations for fermions on a triangular lattice. SDW order on a triangular lattice develops with the momentum ${\bf K} = (4\pi/3,0)$ (for co-planar order in e.g.\ $x-z$ plane).
  We argue that the prefactors in the diagrammatic series for the thermal self-energy depend on the ratio of in-plane and out-of-plane susceptibilities, and by changing this ratio one can control the outcome of the summation of the series. This introduces a control parameter, by which one can vary the strength of the pseudogap behavior. We show that, for a certain value of the control parameter, the system does not develop the pseudogap, despite that self-energy corrections are singular. We note in passing that there is a similarity between this last case in our model and the ``supermetal" scenario for 2D fermions near a single Van-Hove point~\cite{Isobe2019}. In both cases, corrections to fermionic propagator are logarithmically singular, yet the system retains a conventional Fermi liquid behavior.

\begin{figure}[t]
\subfigure[]{\includegraphics[width=0.48\columnwidth]{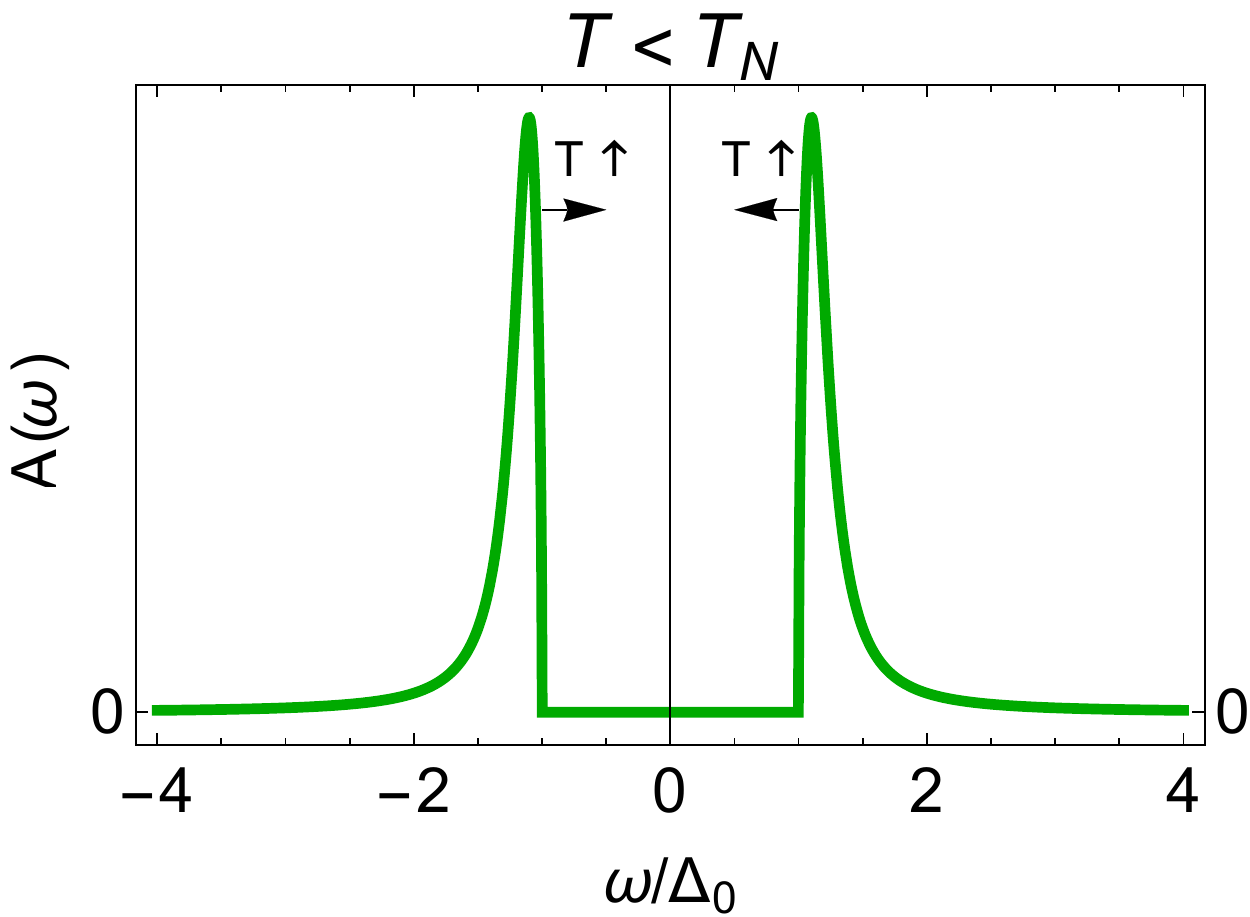}}
\subfigure[]{\includegraphics[width=0.48\columnwidth]{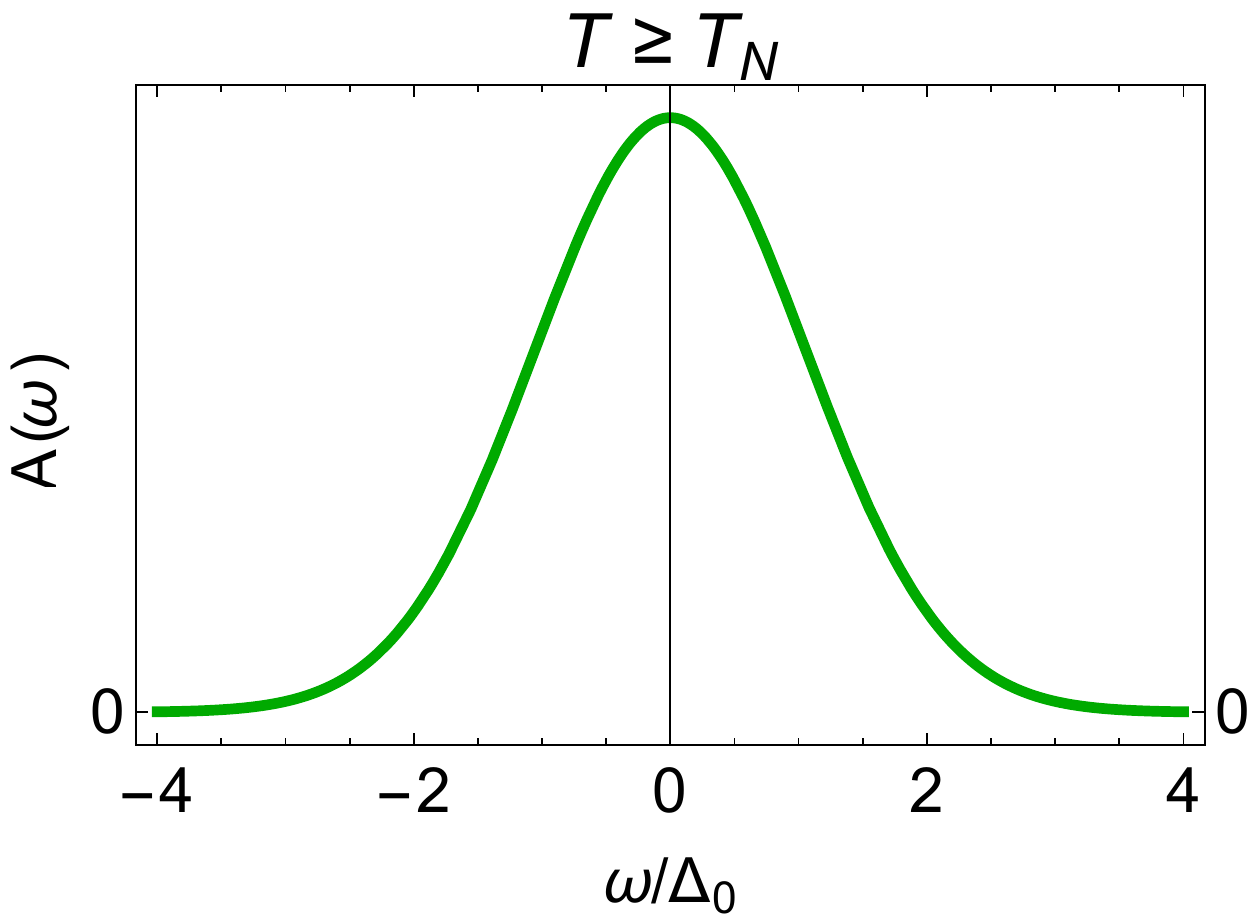}}
\caption{
The evolution of the spectral function  in mean-field approximation, at a hot spot on the Fermi surface.  
(a) In the SDW state, the spectral function has two peaks at energies $\pm \Delta (T)$, where $\Delta (T)$ is proportional to the magnitude of SDW order parameter. 
(b) AT $T = T_N$, the two peaks merge, and at $T > T_N$, the spectral function has a single maximum at $\omega =0$, like in an ordinary metal. The peaks are $\delta-$functional in ``pure" mean-field approximation, but get broadened by regular (i.e., non-logarithmical) thermal and quantum fluctuations. We added a finite broadening phenomenologically to model these effects.}
 \label{fig:summaryMF}
\end{figure}

\begin{figure*}[t]
	\subfigure[]{\includegraphics[width=0.6\columnwidth]{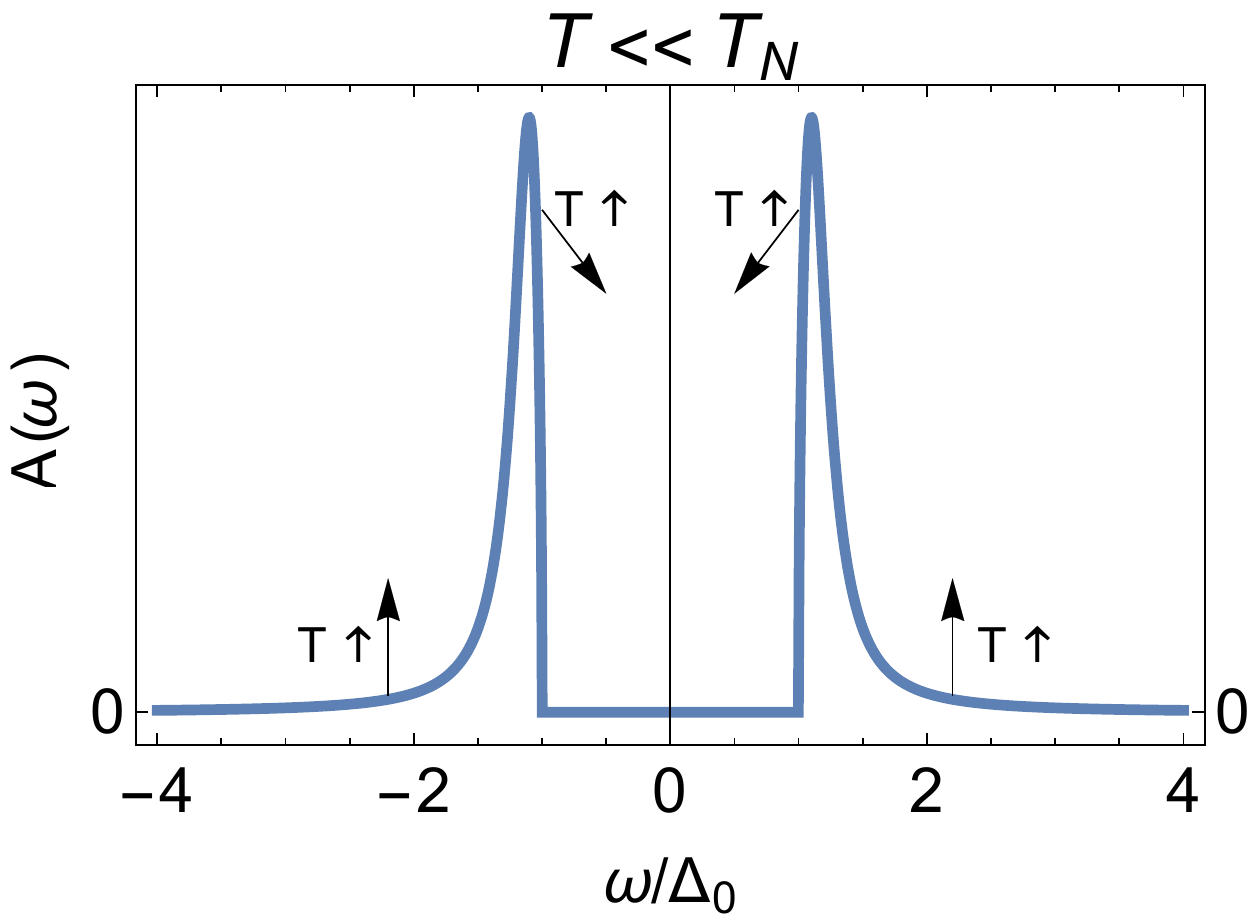}}
	\subfigure[]{\includegraphics[width=0.6\columnwidth]{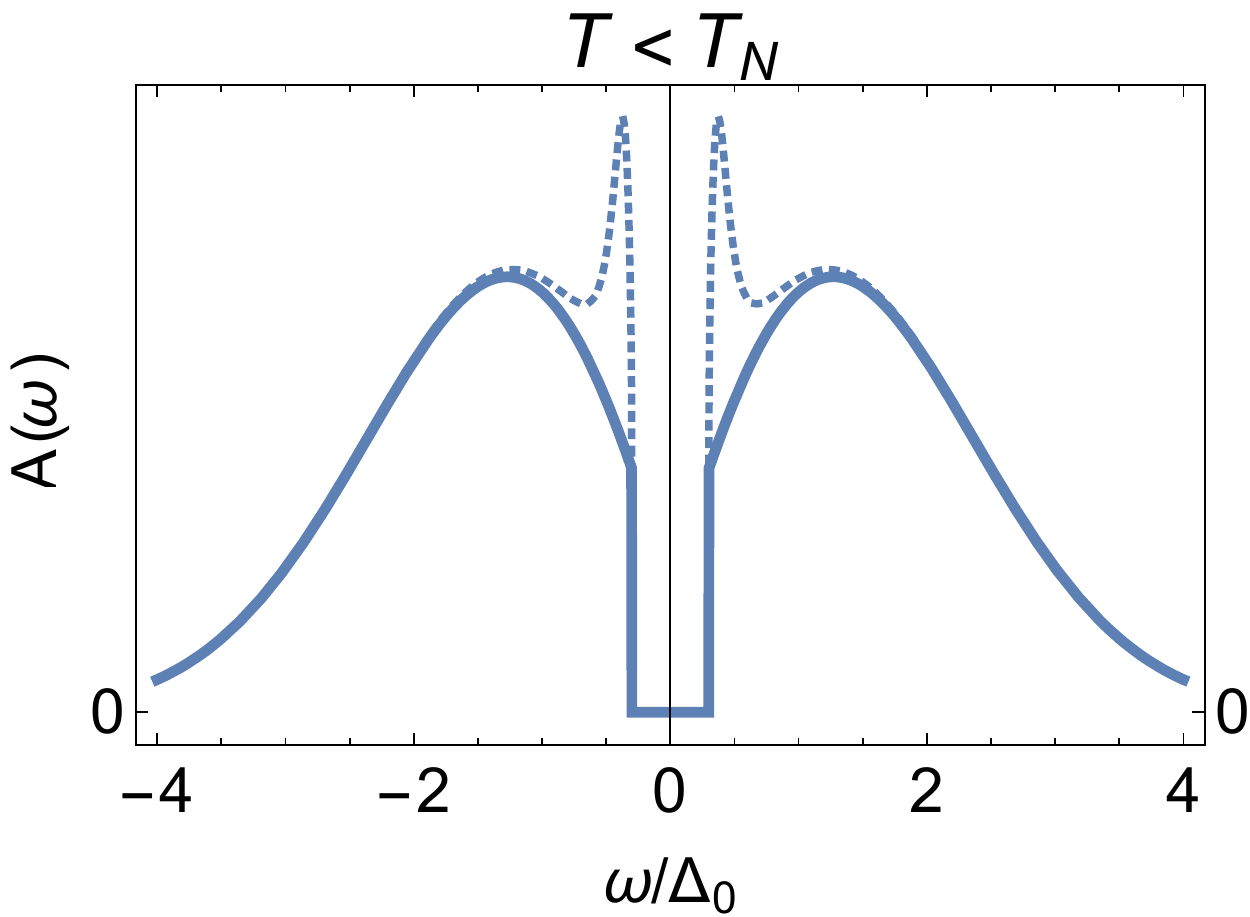}}
	\subfigure[]{\includegraphics[width=0.6\columnwidth]{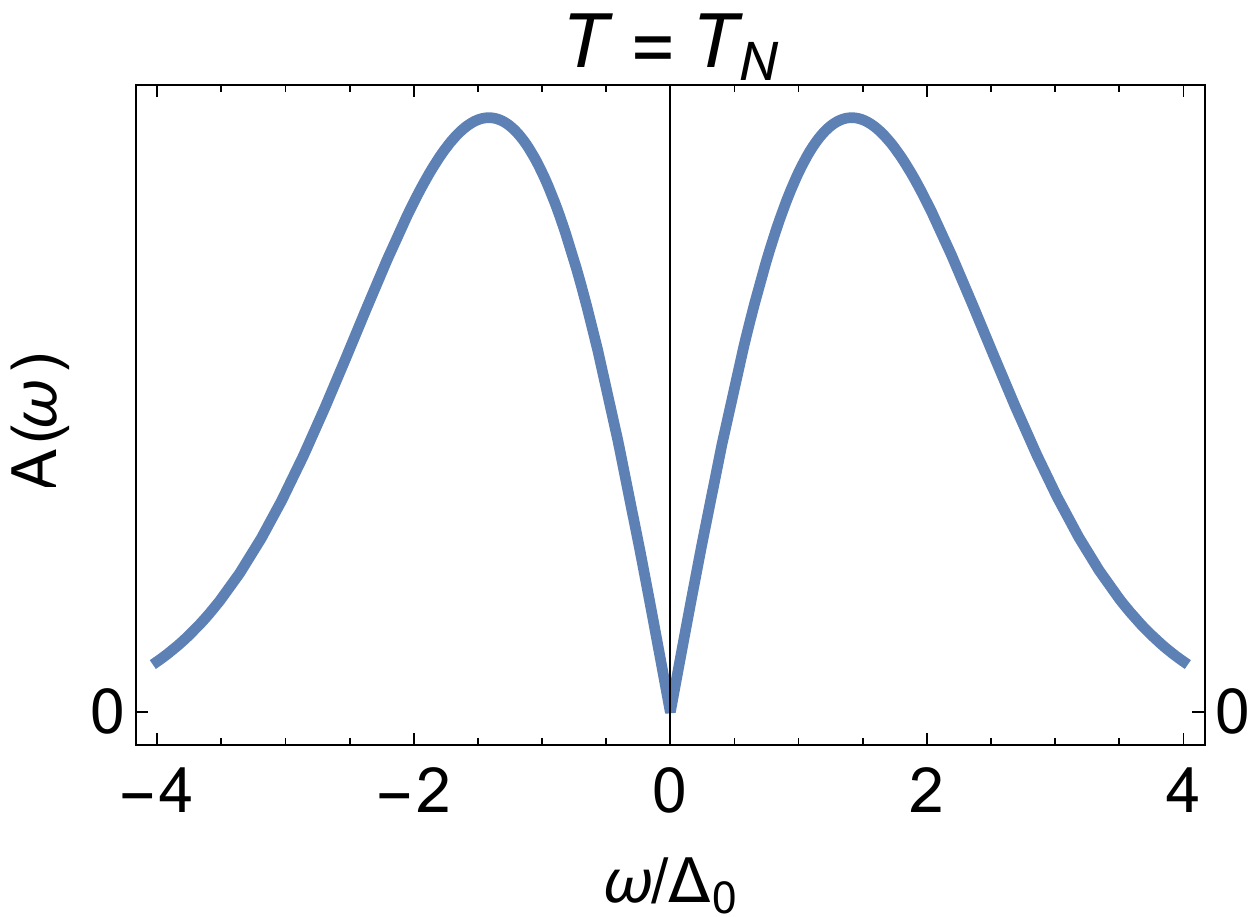}}  \\
	\subfigure[]{\includegraphics[width=0.6\columnwidth]{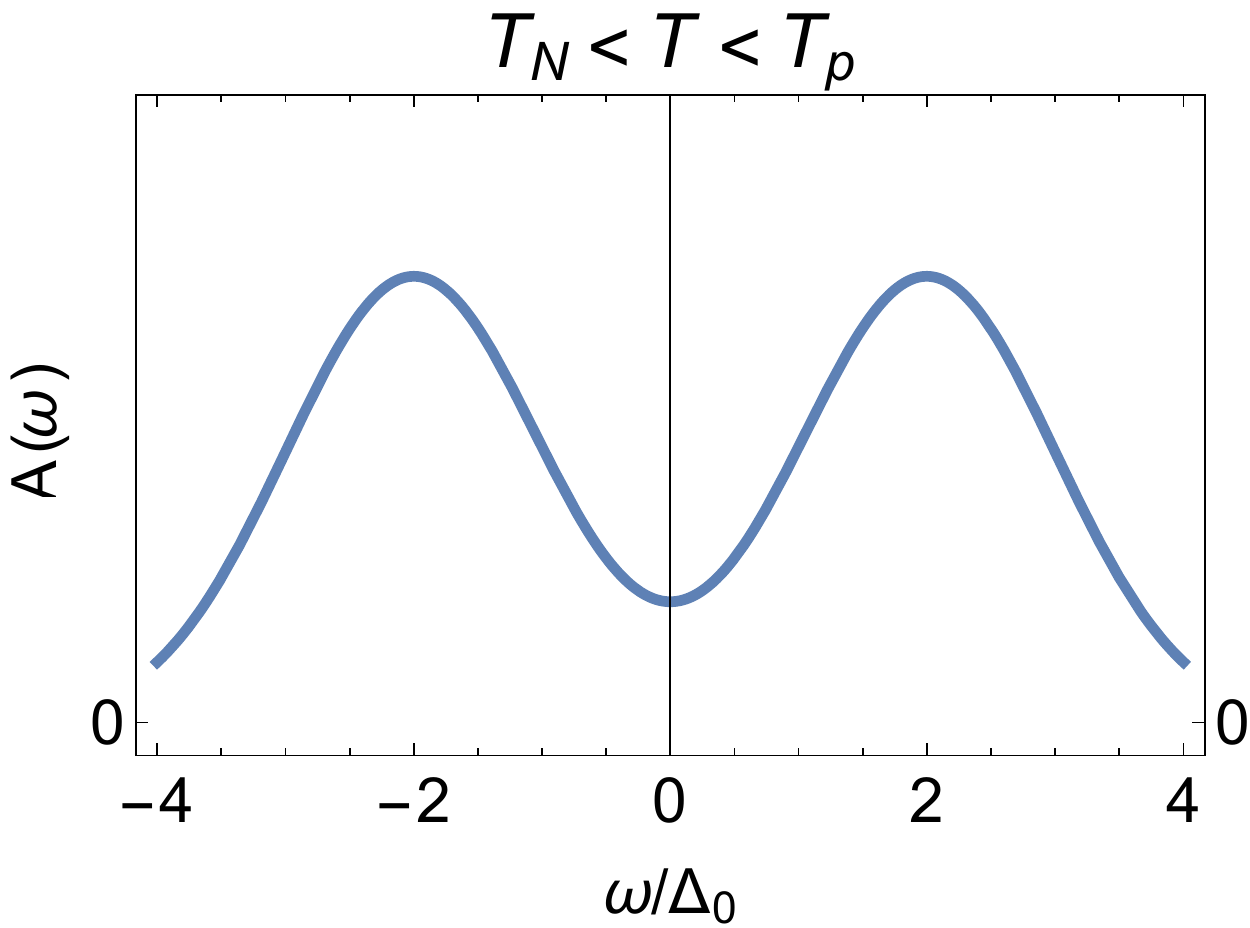}}\quad\quad
	\subfigure[]{\includegraphics[width=0.6\columnwidth]{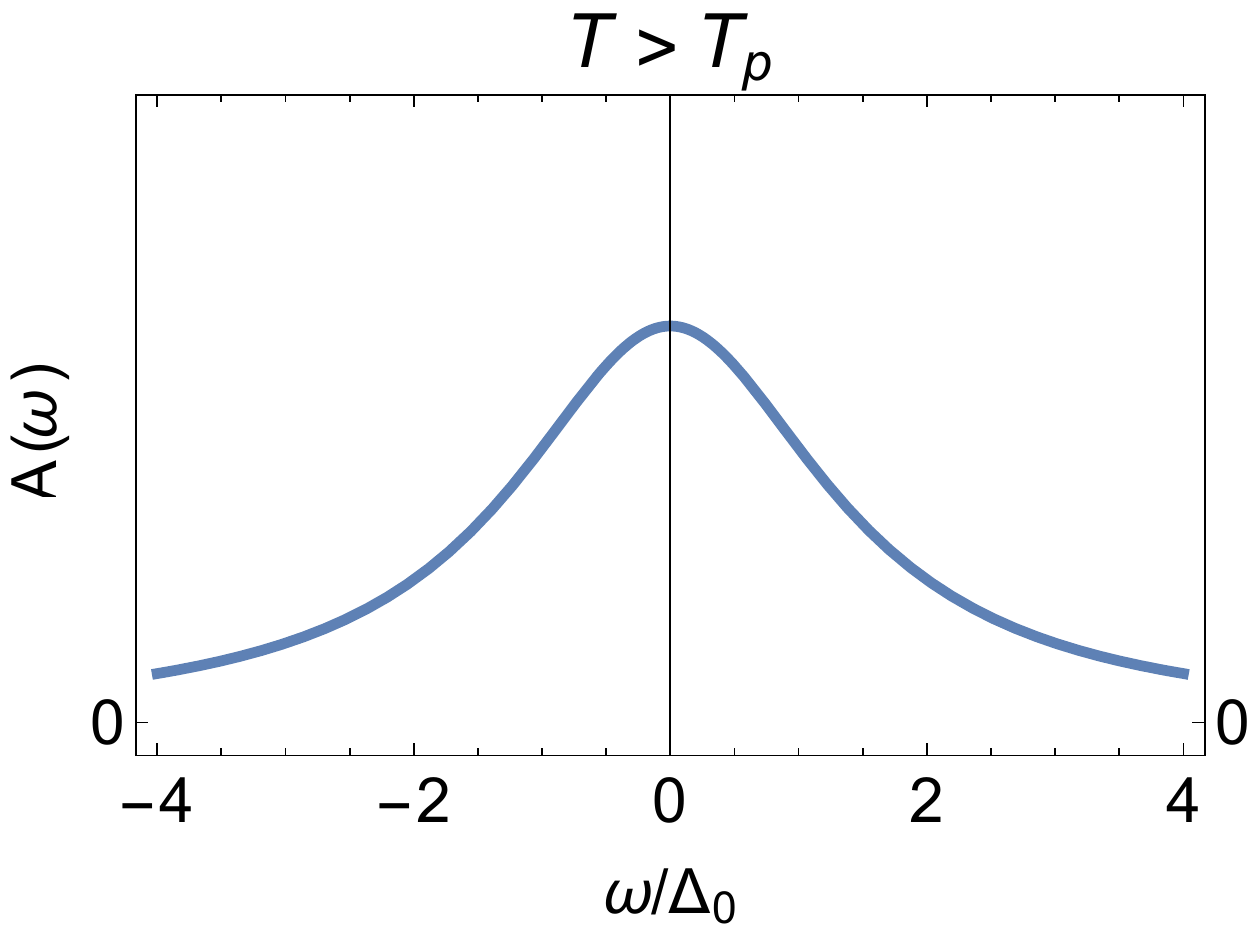}}
\caption{
The sketch of the evolution of the spectral function at a hot spot, when series of logarithmical corrections from thermal fluctuations are included. (a) Deep inside the SDW phase, the spectral function is the same as in mean-field approximation - there are peaks at $\omega = \pm \Delta (0)$. (b) At  $T  \leq  T_N$, the spectral weight vanishes at $|\omega| < \Delta (T)$, like in mean-field, but the spectral function also develops a hump at $|\omega| \sim \Delta (0)$. (c) At $T = T_N$, the true gap vanishes, but the hump remains. (d) At $T \geq T_N$, the spectral function is non-zero at all frequencies, but has a minimum at $\omega =0$ rather than a peak. This has been termed as pseudogap behavior. (e) The conventional metallic behavior is restored only at $T>T_p$ ($T\gg T_N$). The temperature around which the hump vanishes is defined as $T_p$. The solid lines in panel (b) show the result for the spectral function, when only singular thermal self-energy corrections are included.  The dashed lines show the full result, including non-singular self-energy corrections (see the discussion in Sec.~\ref{sec:4-3}).}
\label{fig:summaryPG}
\end{figure*}

In our analysis we primarily focus on the ``hot spots" in momentum space, i.e., on the ${\bf k}$  points, for which ${\bf k}$ and ${\bf k} + {\bf K}$ are both on the Fermi surface without SDW. An SDW order $\Delta (T)$ opens up a spectral gap at these ${\bf k}$. In the mean-field approximation, the spectral function $A_k (\omega)$ at hot spots then has two peaks at $\omega = \pm \Delta (T)$ (see Fig.~\ref{fig:summaryMF}(a)). Within mean-field, the peak position scales with the magnitude of a SDW order and vanishes right at $T_N$, where the order disappears. At higher $T$, the spectral function of a hot fermion is peaked at $\omega =0$, as is expected for a fermion on the Fermi surface in an ordinary paramagnetic metal (see Fig.~\ref{fig:summaryMF}(b)). Thermal fluctuations can change this behavior. For a generic value of our control parameter, $A_k (\omega)$ in a SDW state displays a peak, a dip, and a hump. A peak is at $\Delta(T)$, a hump is near $\Delta (T=0)$, and a dip is in between these two scales (see Fig.~\ref{fig:summaryPG}(b)).  The spectral function almost vanishes below the peak, i.e., a true gap is $\Delta (T)$, like in a mean-field approximation. However, the spectral weight in the peak is reduced compared to a mean-field $A_k (\omega)$, and the difference is transferred into a hump. Above $T_N$, the peak disappears, but the hump survives up to $T = T_p > T_N$.  In between $T_N$ and $T_p$, the spectral function  at a hot spot is non-zero at $\omega =0$, like at ${\bf k}_F$ in a ordinary metal, but the maximum in $A_k (\omega)$ remains at a finite frequency, i.e., the system displays a pseudogap behavior (see Fig.~\ref{fig:summaryPG}(d)). As $T$ increases towards $T_p$, the value of $A_k (0)$ increases, and above $T_p$, the maximum of $A_k (\omega)$  moves to $\omega =0$ (see Fig.~\ref{fig:summaryPG}(e)).

%\begin{widetext}
%\onecolumngrid
%\end{widetext}

For the special value of our control parameter, the spectral function below $T_N$ has only a peak at $\omega = \pm \Delta (T)$, but no hump.  Above $T_N$, the peak disappears, and the spectral function has a single maximum at $\omega =0$, like in ordinary paramagnetic metal.  In this situation, precursor behavior does not develop. Still, even in this case, the  spectral function dressed by thermal SDW fluctuations is quite different from the mean-field $A_k (\omega)$.

For our analysis, we consider the Hubbard model on a triangular lattice, in the large $U$ limit.  At $T=0$, this model displays a co-planar, $120^0$ SDW order with ordering momentum ${\bf K} = (4\pi/3,0)$. Within mean-field, the magnitude of the SDW order is $\Delta(0) = U |\langle\vec{S}\rangle| = U/2$ at half-filling. Quantum fluctuations reduce $|\langle\vec{S}\rangle|$ by about $50\%$~\cite{Jolicoeur1989,Capriotti1999}, but do not destroy the order, nor change that $\Delta (0) \sim U$.  The order gaps fermions at hot spots, and the distance between the conduction and the valence band is $2 U |\langle\vec{S}\rangle|$ (which in the mean-field approximation is the Hubbard $U$).

At a finite $T$, the order is strongly affected by thermal fluctuations. In a 2D system, they destroy long-range order at any finite $T$. In a quasi-2D system, which we consider, the corrections to $|\langle\vec{S}\rangle|$ scale as $(T/J) |\log{\epsilon}|$, where $J = O(t^2/U)$ is the exchange interaction ($t$ is the hopping), and $\epsilon$ measures the deviations from pure two dimensionality. Long-range order get destroyed at $T_N \sim J/|\log{\epsilon}|$.

We will use three simplifications in our analysis. First, we neglect thermal variation of the chemical potential $\mu$.  In principle, $\mu (T)$ has to be computed simultaneously with the fermionic self-energy, from the condition on the total number of fermions. In the SDW state, and in the pseudogap state above $T_N$, $\mu(T)$ by itself evolves with  temperature, and this evolution keeps the position of the hump at distance $O(J)$  from the Fermi energy. As we said, our key goal is to analyze how the pseudogap behavior varies as we change the control parameter. Because the distance between the two humps at positive and negative $\omega$ in Fig. \ref{fig:summaryPG} does not depend on $\mu$, we will not include the thermal evolution of $\mu$ into our analysis and just use the non-pseudogap normal state value for $\mu$. As the consequence of keeping $\mu$ fixed, the spectral function $A_k (\omega)$ at a hot spot is a symmetric function of frequency, and the positions of the peak and the hump in the SDW state and  the hump in the pseudogap state are all set by $U$. Second, we compute the spectral function only at $T < T_N$. This is enough for our purpose. Indeed, it is clear from Fig.~\ref{fig:summaryPG} that when the spectral function retains a hump at $T= T_N-0$, it necessary displays a pseudogap behavior at $T > T_N$.  And, likewise, when the spectral function does not have a hump at  $T= T_N-0$, it does not display a pseudogap behavior at $T > T_N$.
Third, in this work we only consider the renormalizations of the fermionic propagator due to an exchange of transverse spin fluctuations. The self-energy due to an exchange of longitudinal spin fluctuations is non-singular and we neglect it. We caution that this last approximation works well at $T$ substantially smaller than $T_N$, when the transverse fluctuations are gapless, but the longitudinal fluctuations are gapped. At $T \approx T_N$, the gap for longitudinal fluctuations gets smaller, and these fluctuations may enhance the tendency towards pseudogap behavior ~\cite{Schmalian1998,*Schmalian1999}.

Before we proceed, we briefly outline the computational procedure. As we said, it is similar to the eikonal approximation in the scattering theory~\cite{Eikonal}. To our knowledge, its was first applied in the solid state context  in the study of one-dimensional (1D) systems with charge density wave (CDW) fluctuations~\cite{Sadovskii1974,*Sadovskii1974b,*Sadovskii1979,*[{For a detailed discussion of the formalism, see }]SadovskiiBook} (see also Refs.~\cite{McKenzie1996}). The eikonal approximation been applied to cuprates to analyze the  precursors of a collinear $(\pi,\pi)$  SDW state, in the paramagnetic phase ~\cite{Schmalian1998,*Schmalian1999} and in the SDW state~\cite{Sedrakyan2010}.  It has also been used in the calculations of non-analytical corrections to Fermi liquid behavior in a 2D metal~\cite{Efetov2006}. In our analysis, we follow Ref. ~\cite{Sedrakyan2010}, introduce valence and conduction bands in the SDW state, and derive the vertex for the interaction between fermions and magnons (Goldstone modes of the transverse fluctuations of the order parameter). The $120^\circ$ SDW order is co-planar, but not collinear, which implies that it fully breaks the $SU(2)$ spin rotation symmetry. As the consequence, there are three Goldstone modes. The first two are associated with transformations that rotate the plane, where the order sets in. The third one rotates the SDW order parameter within the plane. Accordingly, there are two different spin susceptibilities, $\chi_{\bot}$ and $\chi_{\|}$~\cite{Azaria1990,Chubukov1994}, for out-of-plane and in-plane rotations, respectively.

At a finite temperature, the leading contributions to the renormalization of the  Green's function come from scattering of thermal bosons (in Matsubara formalism, this corresponds to scattering processes with zero transferred bosonic frequency).  These thermal self-energy contributions are logarithmically singular, and scale as powers of $T/J |\log\epsilon|$, where, we remind, $\epsilon$ measures the deviations from pure two dimensionality (see Eq.~\eqref{eq:OneLoop} below). We use $|\log\epsilon|\gg 1$ as a parameter, which allows us to separate singular contributions due to thermal fluctuations from regular contributions of these fluctuations (the terms of order $T/J$, without  $|\log\epsilon|$), and perturbative contributions of quantum fluctuations. The latter are of order one, but are not relevant from physics perspective and can be safely neglected in our analysis of the pseudogap due to precursors to SDW~\cite{Abanov2003} (non-perturbative quantum corrections may be quite relevant~\cite{Sachdev2018}, but this goes beyond the scope of our work). We assume that $T_N/J$ is small and for $n$-loop self-energy keep only the terms of order $(T/J |\log\epsilon|)^n$.  These terms contain both self-energy and vertex corrections, which we put on equal footings.

We show that the combinatoric factor at $n$-loop order, (which we define as $\mathcal{C}_n$ in the text) scales factorially with $n$, and its magnitude depends on the ratio $\chi_{\|}/\chi_{\bot}$. We sum the contributions from all orders and obtain the full self-energy. We then convert from Matsubara to real axis and obtain the spectral function $A_k (\omega)$.

The rest of the paper is organized as follows. In Sec.~\ref{sec:2-1} we introduce the Hamiltonian, discuss mean-field solution, and obtain the dynamical magnetic susceptibility associated with the Goldstone modes, the effective 4-fermion interaction mediated by magnons, and the magnon-fermion vertex function.
In Sec.~\ref{sec:3}, we obtain and sum up the series of leading logarithmical diagrams for the fermionic Green's function.
In Sec.~\ref{sec:4} we obtain the spectral function for different ratios of $\chi_{\|}/\chi_{\bot}$. This is the main result of the paper. In Sec.~\ref{sec:discussions} we discuss the results and summarize our findings.

%========================================================
%========================================================
\section{The model}
%========================================================
\subsection{The Hamiltonian and the mean field solution}\label{sec:2-1}
The point of departure for our analysis is the one band Hubbard model for spin $1/2$ fermions on a triangular lattice
\begin{align}
\label{eq:Hubbard}
\mathcal{H}=&-\sum_{\langle i, j \rangle,\sigma}t_{i,j}(\cd_{i,\sigma}c_{j,\sigma}+\cd_{j,\sigma}c_{i,\sigma})-\mu\sum_{i}\cd_i c_i\non\\
&+U\sum_i n_{i\uparrow}n_{i\downarrow}
\end{align}
Without loss of generality, we restrict the hopping to nearest neighbors.

We take as an input the fact that the ground state of the model for large $U$ is the $120^{\circ}$ co-planar SDW order. Without loss the generality, we set the global coordinate such that the coplanar order is in the x-z plane, and $\langle S_i^z\rangle + i \langle S_i^x\rangle = \bar S \exp (\pm i {\bf K}\cdot {\bf R}_i)$, where ${\bf K}=(4\pi/3,0)$ and $\bar S$ is the magnitude of average magnetization ($\bar S\approx 1/2$ at mean field level near half-filling and in the large $U$ limit).
The $\pm$ sign in the exponent determines whether the ``direction" of the spiral, i.e., whether the order is $+120^{\circ}$ or $-120^{\circ}$.
Without loss of generality we consider the $+$ case in the rest of the paper.
We introduce the rotating reference frame~\cite{Tremblay1995}, in which all spins are along the same direction $\tilde z_i$. The transformation of fermionic operators $c_i$ to the new basis is given by $c_i=T_i \tilde c_i$, where $T_i=\exp(-i {\bf K}\cdot {\bf R}_i \sigma_y/2)$. One can straightforwardly verify that in the new coordinate frame $\langle \tilde {\bf S}_i\rangle = (1/2) \langle{\tilde c}^\dagger_{i,\alpha} {\bf \sigma}_{\alpha \beta} {\tilde c}_{i,\beta}\rangle = \{0,0,\bar S\}$, i.e. the original $120^{\circ}$  SDW order becomes  ferromagnetic. The Hubbard Hamiltonian in the rotating reference frame takes the form
\begin{align}
\label{eq:Hubbard1}
\mathcal{H}=&-\sum_{\langle i, j \rangle}t_{i,j}(\tilde c^{\dagger}_{i}T_{i,j}\tilde c_{j}+\tilde c^{\dagger}_{j}T_{j,i}\tilde c_{i})-\mu\sum_{i}\tilde c^{\dg}_i \tilde c_i\non\\
&+U\sum_i \tilde c^{\dagger}_{i,+}\tilde c_{i,+}\tilde c^{\dagger}_{i,-}\tilde c_{i,-},
\end{align}
where $T_{i,j}=T_i^{\dagger}T_j$. In explicit form, we have
\begin{align}\label{eq:Tmatrix}
T_{i,j}=
\begin{pmatrix}
\cos ({\bf K}\cdot {\bf R}_{ij}/2) & \sin  ({\bf K}\cdot {\bf R}_{ij}/2)\\
-\sin  ({\bf K}\cdot {\bf R}_{ij}/2) & \cos ({\bf K}\cdot {\bf R}_{ij}/2)
\end{pmatrix},
\end{align}
where ${\bf R}_{ij}={\bf R}_i- {\bf R}_j$.

The quadratic Hamiltonian Eq.~\eqref{eq:Hubbard1} in momentum space is:
\begin{widetext}
\begin{align}
\mathcal{H}_{quad}=\sum_{k}
\begin{pmatrix}
{\tilde c}^{\dg}_{k,+} & {\tilde c}^{\dg}_{k,-}
\end{pmatrix}
\begin{pmatrix}
(\epsilon_{k+}+\epsilon_{k-})/2-\mu & i(\epsilon_{k+}-\epsilon_{k-})/2\\
-i(\epsilon_{k+}-\epsilon_{k-})/2 & (\epsilon_{k+}+\epsilon_{k-})/2-\mu
\end{pmatrix}
\begin{pmatrix}
{\tilde c}_{k,+} \\
 {\tilde c}_{k,-}
\end{pmatrix},
\end{align}
\end{widetext}
where $\epsilon_{k+} = \epsilon_{k+K/2}$, $\epsilon_{k-} = \epsilon_{k-K/2}$ and $\epsilon_k = -t\sum_{\langle i, j\rangle}\exp (i k\cdot r_{ij}) = -2t(\cos k_x+4\cos k_x/2\cos \sqrt{3}k_y/2).$
To simplify notations, we  define $\omega_{k}=\frac{1}{2}(\epsilon_{k+}+\epsilon_{k-})$, $\eta_k=\frac{1}{2}(\epsilon_{k+}-\epsilon_{k-})$.
In these notations,
\begin{align}
\mathcal{H}_{quad}=\sum_{k}
\begin{pmatrix}
{\tilde c}^{\dg}_{k,+} & {\tilde c}^{\dg}_{k,-}
\end{pmatrix}
\begin{pmatrix}
\omega_k-\mu & i\eta_k\\
-i \eta_k  & \omega_k-\mu
\end{pmatrix}
\begin{pmatrix}
{\tilde c}_{k,+} \\
 {\tilde c}_{k,-}
\end{pmatrix}.
\end{align}

\subsubsection{The mean-field approximation}

In the SDW state, $\langle {\tilde c}^{\dg}_{i,+}{\tilde c}_{i,+}\rangle=-\langle {\tilde c}^{\dg}_{i,-}{\tilde c}_{i,-}\rangle=\bar S$. The interaction term in mean field approximation reduces to
\begin{align}
\mathcal{H}_{int}&=U\sum_i \tilde c^{\dagger}_{i,+}\tilde c_{i,+}\tilde c^{\dagger}_{i,-}\tilde c_{i,-}\non\\
&\rightarrow\Delta  c^{\dagger}_{i,-}\tilde c_{i,-}- \Delta c^{\dagger}_{i,+}\tilde c_{i,+},
\end{align}
where $\Delta=U \bar S$. The mean field Hamiltonian becomes
\begin{align}\label{eq:Hmf}
\mathcal{H}_{MF}=
\begin{pmatrix}
{\tilde c}^{\dg}_{k,+} & {\tilde c}^{\dg}_{k,-}
\end{pmatrix}
\begin{pmatrix}
\omega_k-\mu-\Delta & i\,\eta_k\\
-i\,\eta_k & \omega_k-\mu+\Delta
\end{pmatrix}
\begin{pmatrix}
{\tilde c}_{k,+} \\
 {\tilde c}_{k,-}
\end{pmatrix}.
\end{align}
This Hamiltonian can be diagonalized by intoducing the operators of canonical modes $\Gamma_k=\{\gamma^c_k,\gamma^v_k\}^T$ via $\Gamma_k=V^\dagger_k \tilde c_k$, where the unitary matrix $V_k$ is
\begin{align}
V_k=
\begin{pmatrix}
\cos \phi_k & i\sin\phi_k\\
i\sin\phi_k & \cos \phi_k
\end{pmatrix},
\end{align}
and
$\cos\phi_k=\sqrt{\frac{1}{2}\big(1-\frac{\Delta}{\sqrt{\Delta^2+\eta_k^2}}\big)}$, $\sin\phi_k=-\sqrt{\frac{1}{2}\big(1+\frac{\Delta}{\sqrt{\Delta^2+\eta_k^2}}\big)}$.
The mean-field Hamiltonian in terms of canonical modes  is
$\mathcal{H}_{MF}=\Gamma_k^{\dg}\Lambda_k\Gamma_k$.
where $\Lambda_k=V_k^{\dg}H_{MF}V_k$ is diagonal. In explicit form
\begin{align}
\mathcal{H}_{MF}=\sum_{k} (E^c_k \gamma^{c\dagger}_k\gamma^c_k+E^v_k \gamma^{v\dagger}_k\gamma^v_k),
\end{align}
and $E^{c}_k=\omega_k+\sqrt{\Delta^2+\eta_k^2}$, $E^{v}_k=\omega_k-\sqrt{\Delta^2+\eta_k^2}$.

A comment is in order here. The expressions for $E^{c}_k$ and $E^{v}_k$ are in the rotated coordinate frame. At $\Delta =0$, $E^{c,v}_k = \epsilon_{\pm} = \epsilon_{{\bf k} \pm {\bf K}/2}$.  A hot spot location ${\bf k}_{hs}$  in the rotated coordinate frame is defined as a point for which $\epsilon_{{\bf k}_{hs}+{\bf K}/2} = \epsilon_{{\bf k}_{hs}-{\bf K}/2} =0$, hence  $E^{c}_{k_{hs}} = E^{v}_{k_{hs}} =0$.   However, once we shift ${\bf k}_{hs}$ by ${\bf K}$, we find that  $E^{c,v}_{k_{hs}+{\bf K}}$  is either zero or $\epsilon_{{\bf k}_{hs}+3{\bf K}/2}$. The latter is numerically small at half-filling but strictly vanishes only for a certain hole doping.
In the SDW state we then have $|E^{c,v}_{k_{hs}}| = \Delta$, but $|E^{c,v}_{k_{hs}+{\bf K}}|$ is not exactly $\Delta$, with some small correction at order $t$.
 Below we neglect this complication and approximate $|E^{c,v}_{k_{hs}+{\bf K}}|$ by $\Delta$.

\begin{figure}[bth!]
\centering
\subfigure[]{\includegraphics[width=0.45\columnwidth]{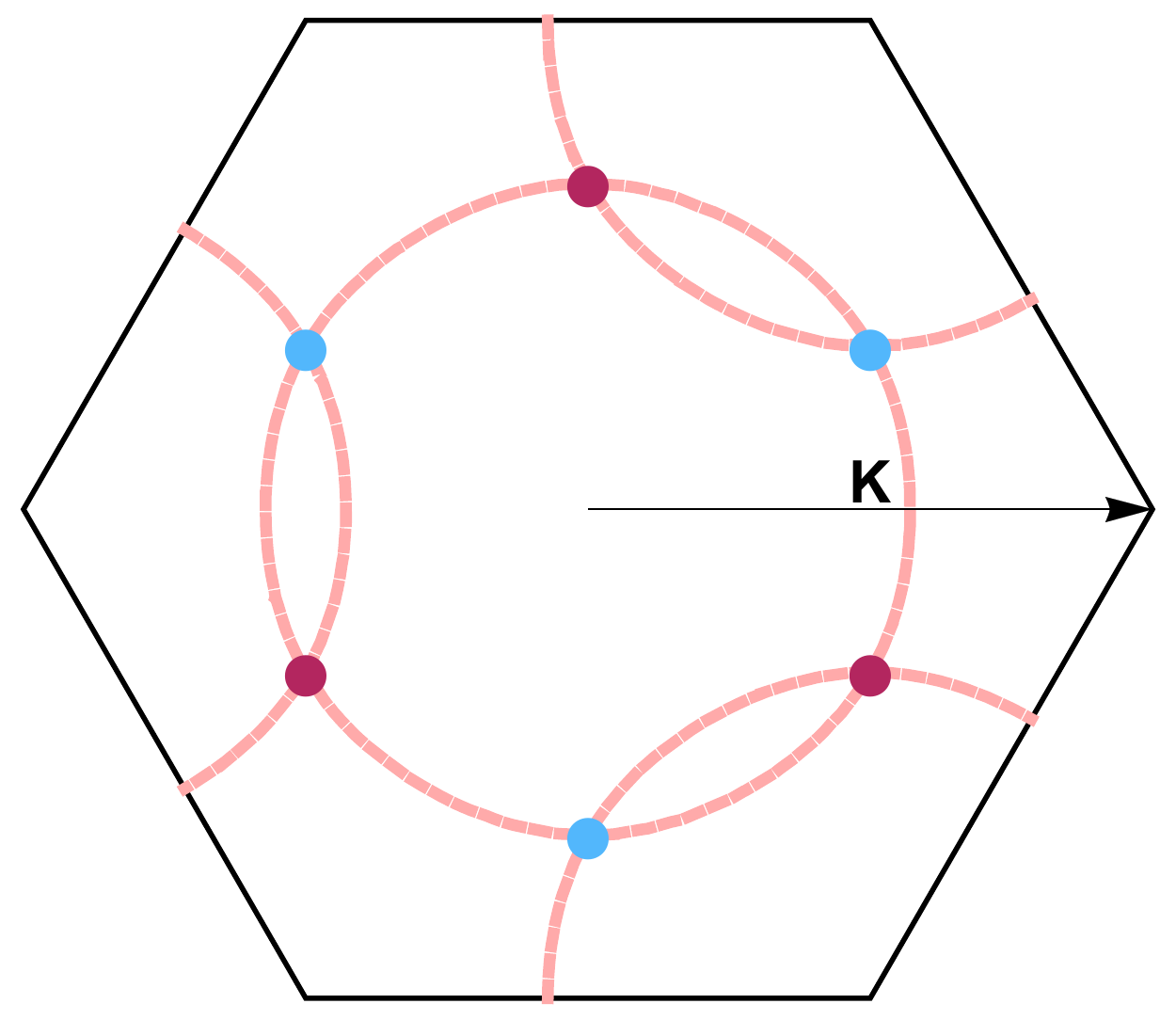}}\quad
\subfigure[]{\includegraphics[width=0.45\columnwidth]{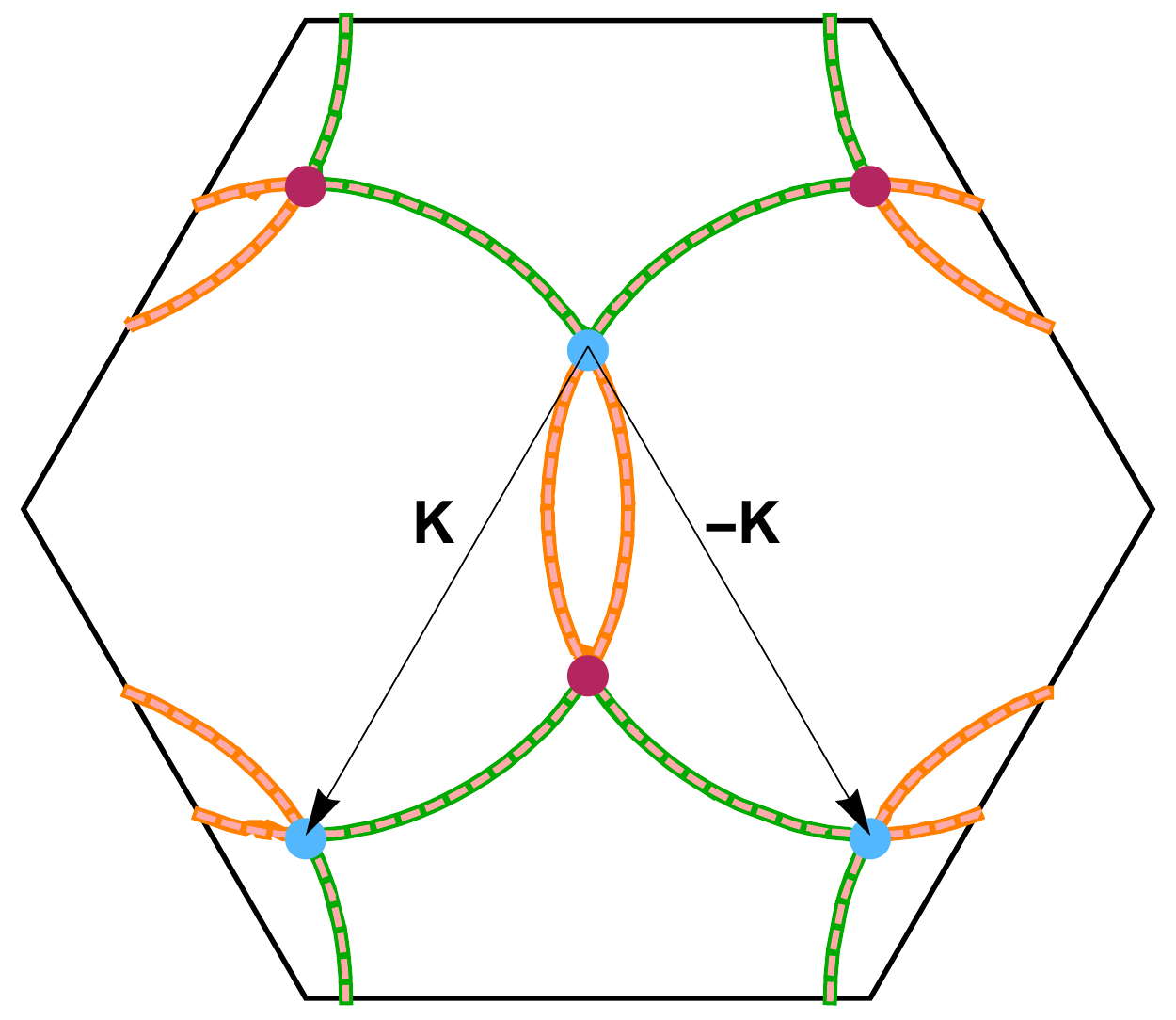}}
\subfigure[]{\includegraphics[width=0.45\columnwidth]{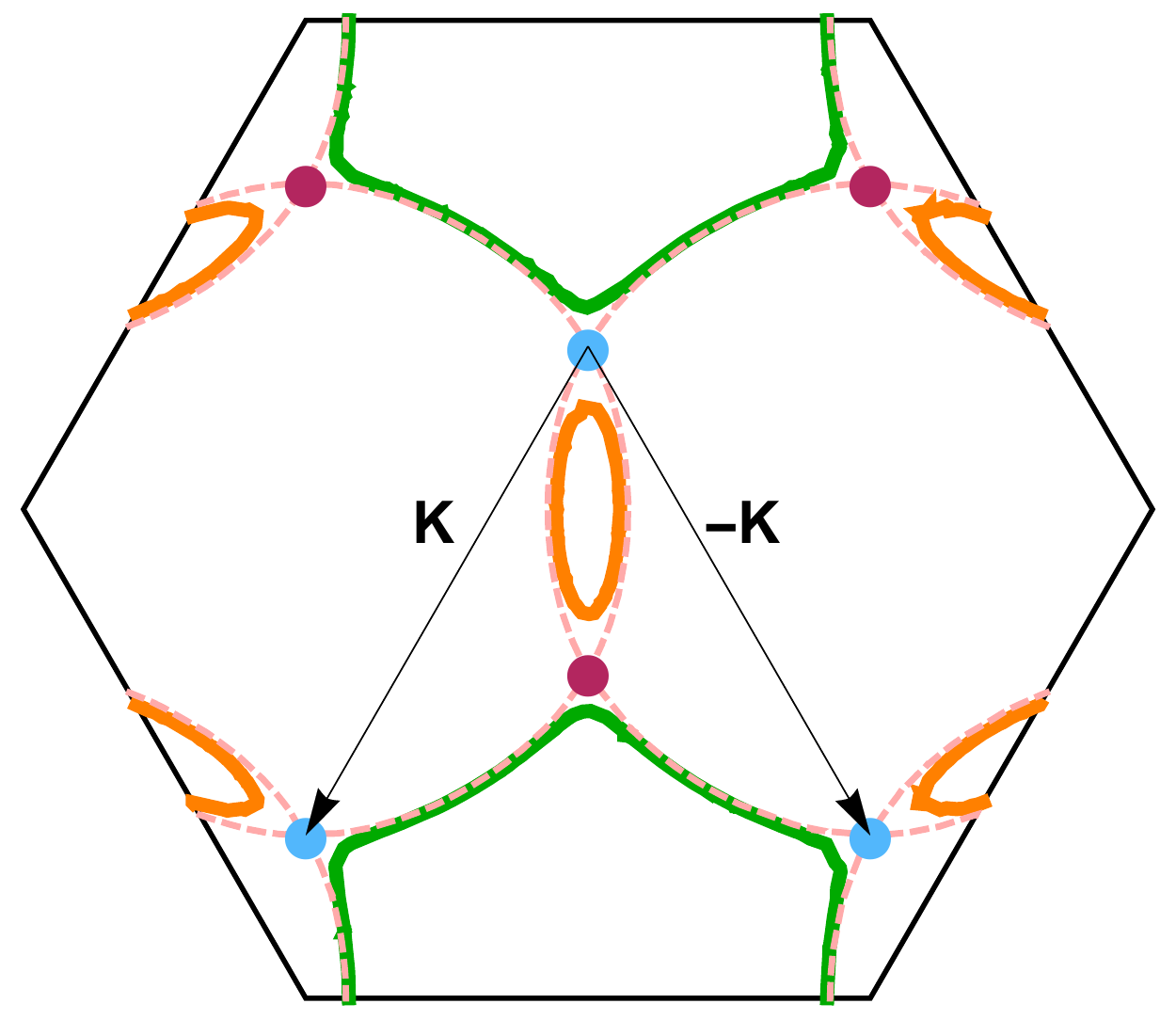}}\quad
\subfigure[]{\includegraphics[width=0.45\columnwidth]{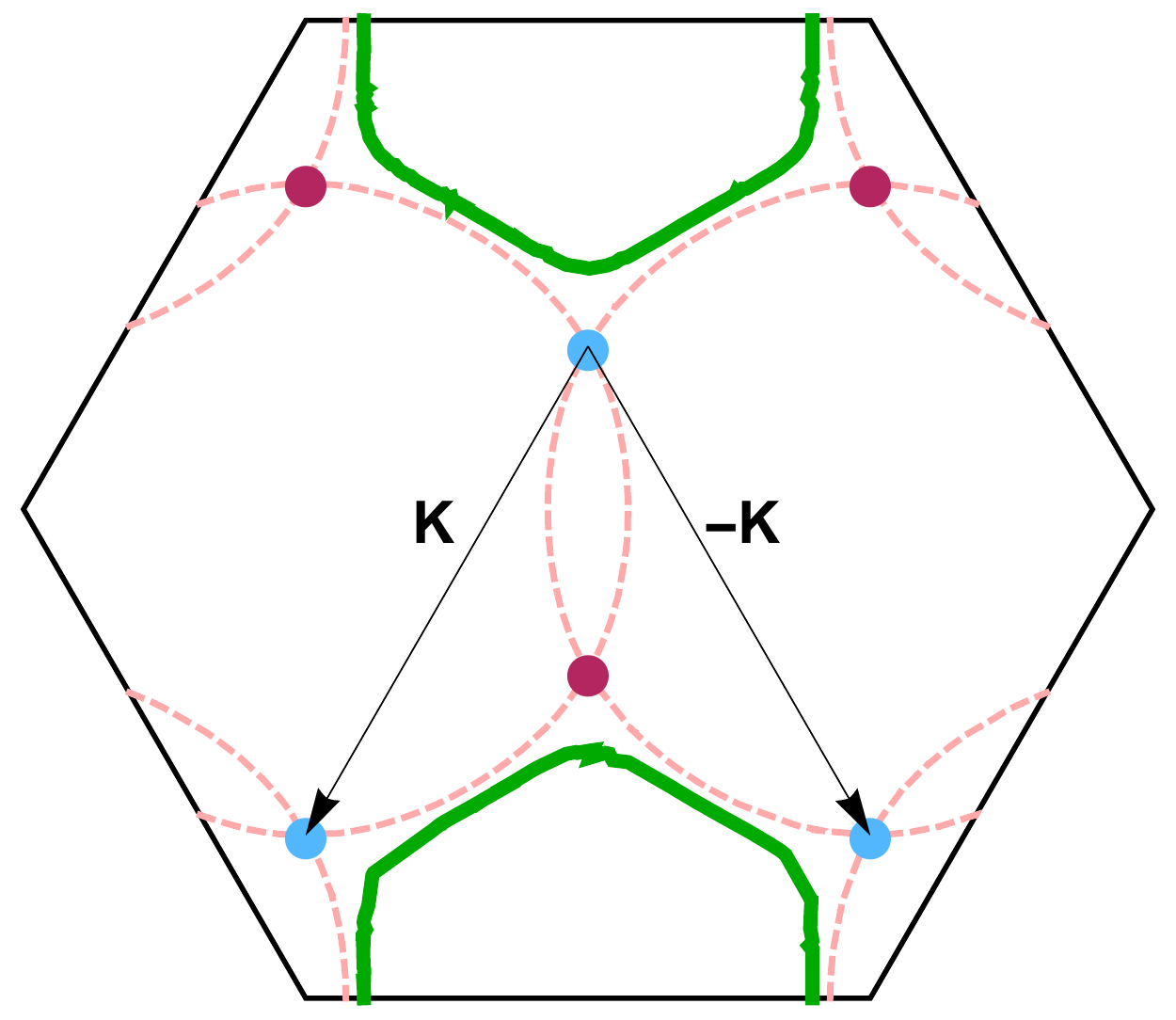}}
\caption{The evolution of Fermi surface at $T+0$ as the SDW order $\Delta$ develops upon increasing of the Hubbard $U$. For definiteness, we consider the case of weak hole doping. (a): The Fermi surface at $U=U_c,\,\Delta=0^{+}$ in the original (not rotated) coordinate frame. The Fermi surface for one spin component is shifted by ${\bf K}$ compared to the Fermi surface for the other spin component. (b)-(d): The evolution of the Fermi surface in the rotated (spin-dependent) coordinate frame. The Fermi surfaces are shifted by ${\bf K}/2$ compared to those in panel (a) (see Eq.~\eqref{eq:Tmatrix}). The \textcolor[rgb]{0,0.3,1}{blue} and \textcolor[rgb]{1,0,0.8}{red} dots mark the hot spots -- the points where the two Fermi surfaces cross at $\Delta=0^{+}$. The three hot spots in \textcolor[rgb]{0,0.3,1}{blue} (\textcolor[rgb]{1,0,0.8}{red}) are connected by the wave vector $\pm K$. Panel (b) -- Fermi surfaces at $U=U_c,\,\Delta=0^{+}$, panels  (c) and (d) -- Fermi surfaces at $U>U_c, \,\Delta>0$. Both electron (orange line) and hole (green line) pockets shrink as $\Delta$ increases.}
 \label{fig:FermiSurface}
\end{figure}

The chemical potential $\mu$ and the SDW order parameter $\Delta$ should be obtained self-consistently as a function of the interaction strength $U$.
At small $U/t$, self-consistent analysis yields $\Delta =0$, i.e., a paramagnetic Fermi liquid state with large Fermi surface remains stable down to $T=0$.  This is similar to the case of the Hubbard model on a square lattice with both nearest and next nearest neighbor hopping. Once the interaction exceeds a threshold, $U> U_c$, the SDW order develops. This changes the Fermi surface topology to a set of small electron and hole pockets. The sizes of electron and hole pockets shrinks as $U$ increases. At half-filling,   both electron and hole pockets vanish the large $U$ limit, i.e., all excitations are gapped:  there is a filled valence band with energy $E^{v}_k=\omega_k-\sqrt{\Delta^2+\eta_k^2}$ and an empty conduction band with energy $E^{c}_k=\omega_k+\sqrt{\Delta^2+\eta_k^2}$. Such a state can be adiabatically connected to a Mott insulator, which, strictly speaking, does not require magnetic order. Away from the half-filling, the size of the remaining electron and hole pockets is determined by the Luttinger theorem for a SDW state~\cite{Altshuler1998}. We show the evolution of Fermi surface geometry with increasing  $U$ in Fig.~\ref{fig:FermiSurface}.

Across the transition at $U_c$, the spectral function at the ``hot spot" $A_{c,v}(k_{hs},\omega)=-\frac{1}{\pi}\text{Im}\, G_{c,v}(k_{hs},\omega)$ changes qualitatively (see Fig.~\ref{fig:summaryMF}).
At $\Delta =0$, $A_{c}(k_{hs},\omega) = A_{v}(k_{hs},\omega) = 1/(\omega + i\delta)$, i.e., the spectral function is strongly peaked at $\omega =0$.
At a finite $\Delta$, $G^{(0)}_{c,v}(k_{hs},\omega)=\frac{1}{\omega + i \delta \mp \Delta}$, and the maximum shifts to $\omega = \pm \Delta$.  In the large $U$ limit, $\Delta=U\bar S$, i.e., the distance between the two peaks is Hubbard $U$, like in a Mott insulator.

%========================================================
\subsection{Magnon-fermion interaction}\label{sec:2-2}
To obtain the Green's function renormalized  by thermal fluctuations in the SDW state, one should first find the effective magnon-fermion interaction vertex.

The magnon propagator can be obtained by either the linear spin wave analysis~\cite{Chubukov1994} or by evaluating the spin susceptibility within the generalized RPA framework in the large $U$ limit~\cite{Tremblay1995}. We  present the  details of the spin wave analysis in Appendix~\ref{app:Goldstone} and here list the results and use physical arguments to rationalize them.

We set the $120^{\circ}$ coplanar order to be in the $x-z$ plane and move to rotating coordinate frame, where the order becomes ferromagnetic.

\begin{figure}[t]
\centering
\subfigure[]{\includegraphics[width=0.4\columnwidth]{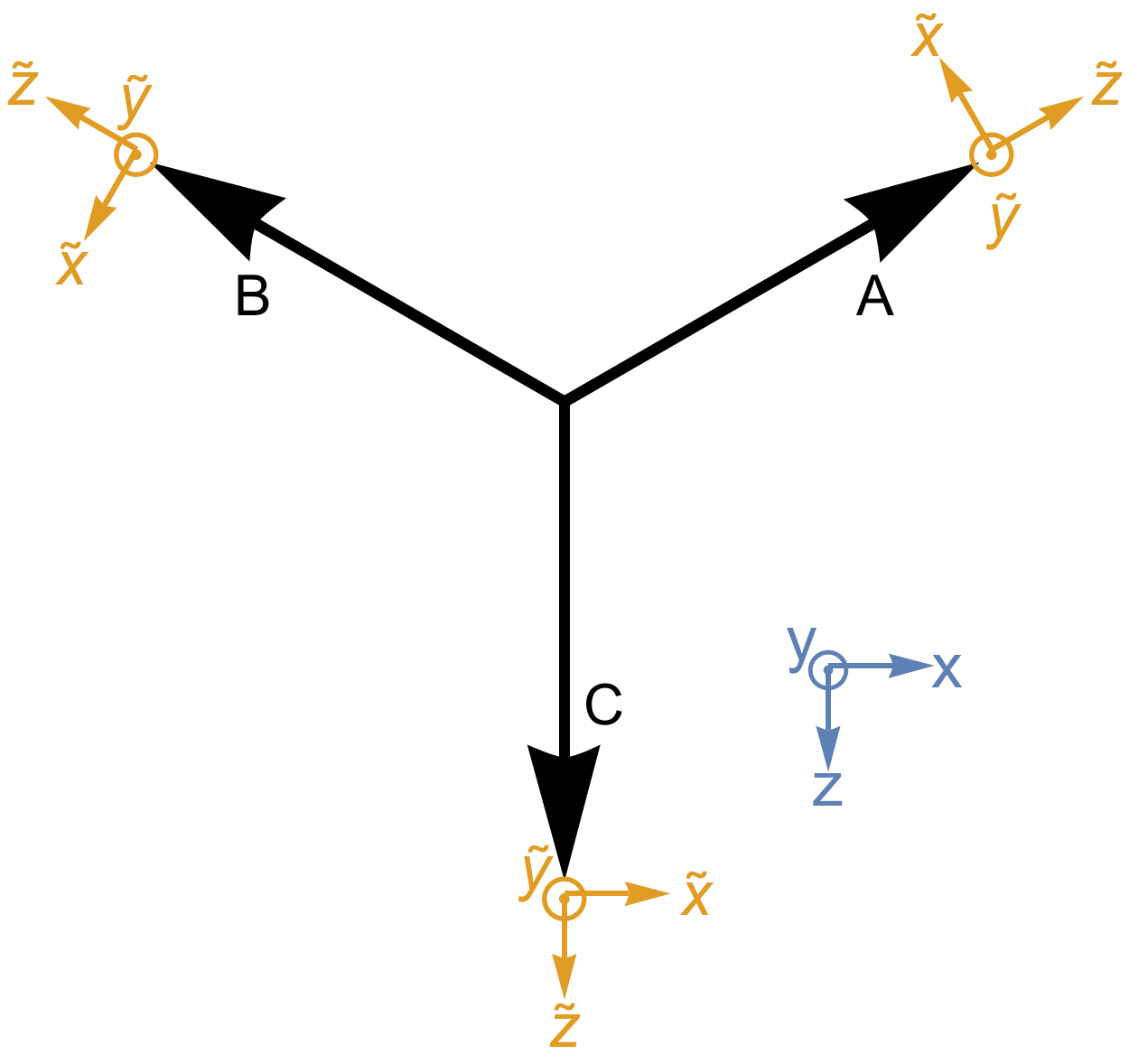}}\quad
\subfigure[]{\includegraphics[width=0.4\columnwidth]{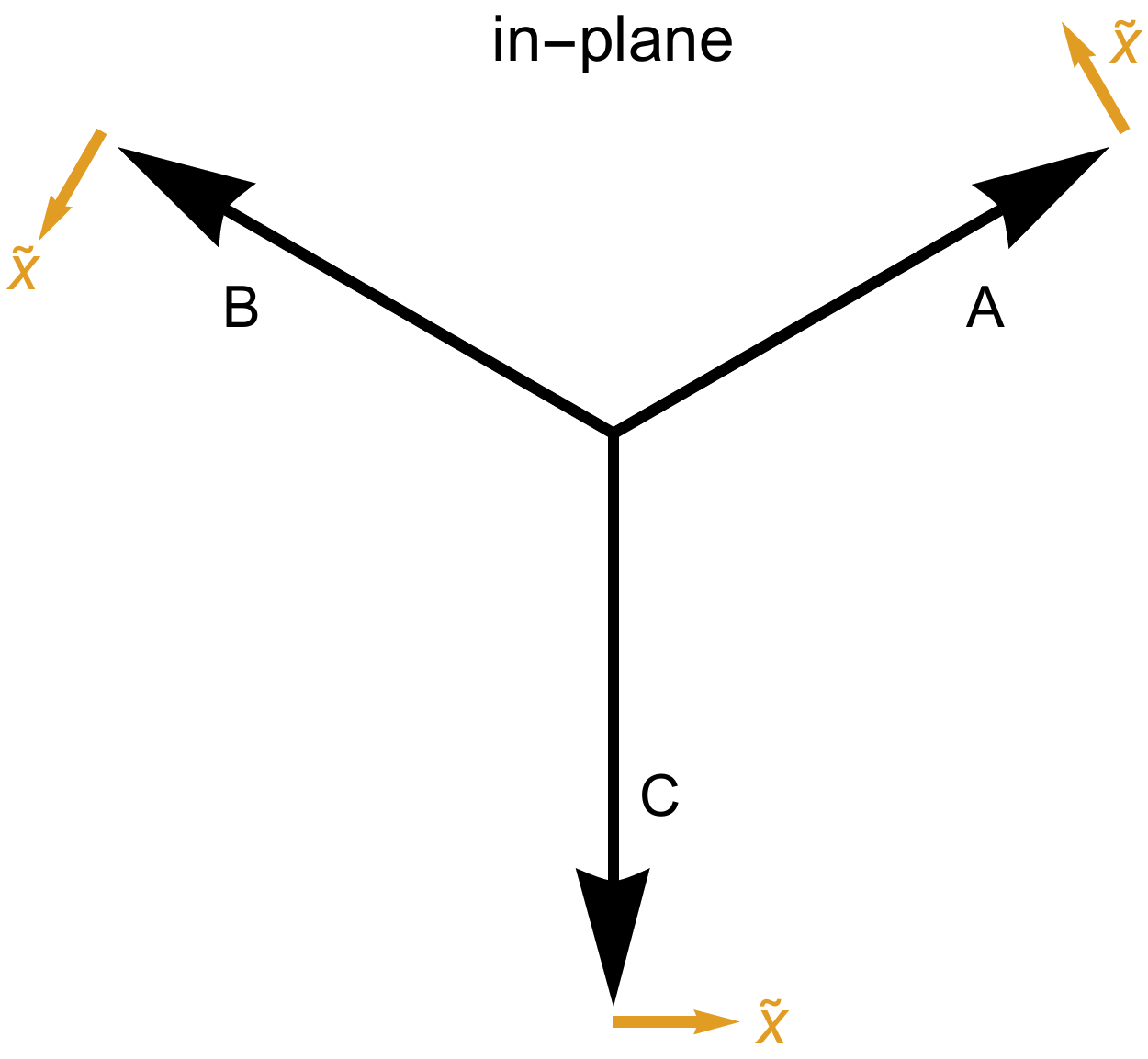}}
\subfigure[]{\includegraphics[width=0.4\columnwidth]{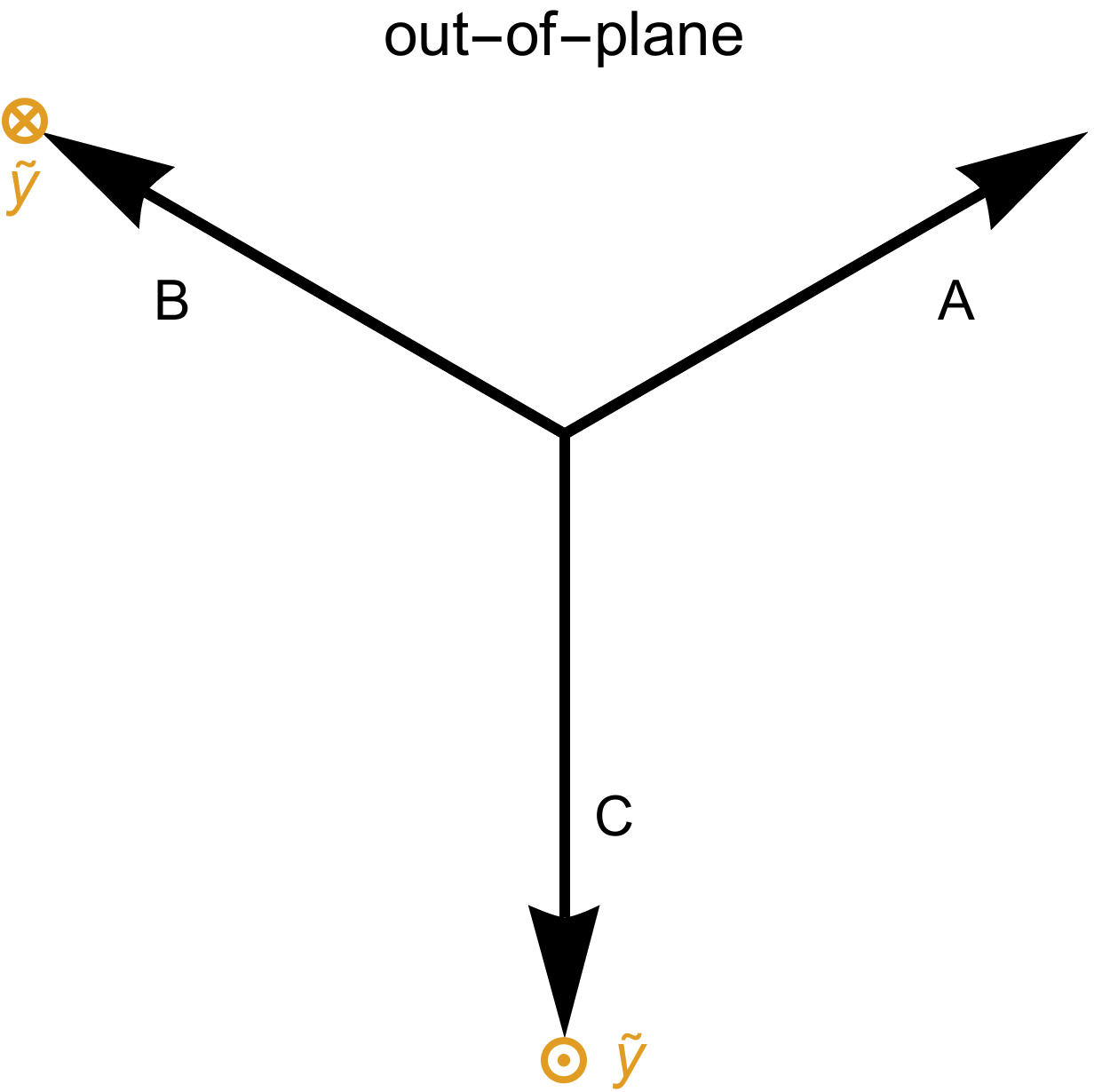}}\quad
\subfigure[]{\includegraphics[width=0.4\columnwidth]{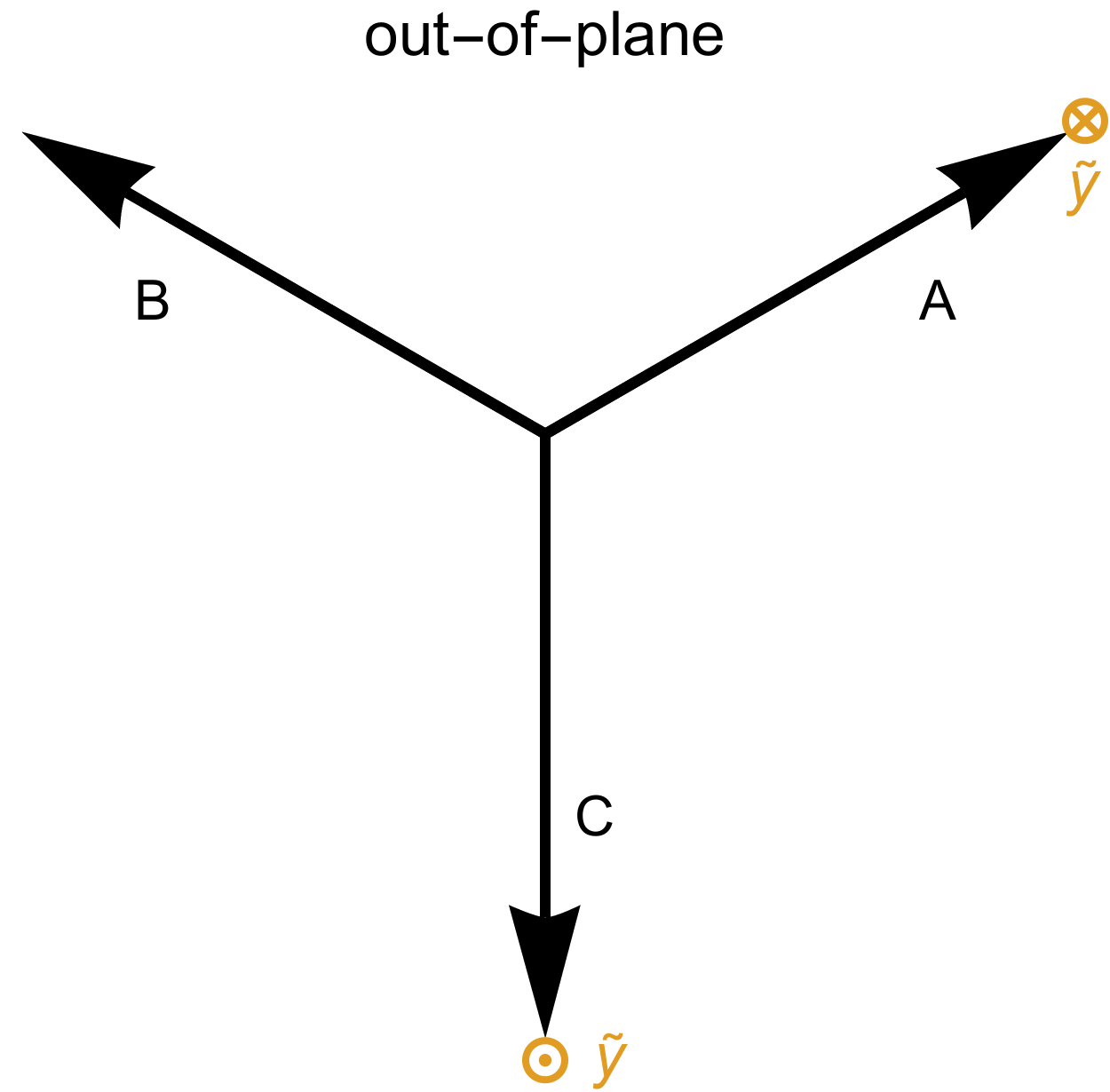}}
\caption{
(a) Magnetic order (black arrow) on three sublattices $A,\,B,\,C$. Blue and orange arrows indicate, respectively, global and local coordinates in spin space. (b-d) Momentum and spin components for the three Goldstone modes [see Eq.~\eqref{eq:Magnon}]. The in-plane Goldstone mode in (b) is described by the pole in $\chi^{\tilde x\tilde x}(\Gamma)$, and the linear combinations of the out-of-plane modes in (c) and (d) are described by the poles in $\chi^{\tilde y\tilde y}(\pm K)$. \label{fig:SpinWaveCo}}
\end{figure}

The the local coordinates for the $A,\,B,\,C$ sublattices are shown in Fig.~\ref{fig:SpinWaveCo}(a). A straightforward symmetry analysis shows that in the SDW state there should be three gapless Goldstone modes, one associated with in-plane fluctuations, and two associated with out-of-plane fluctuations~\cite{Azaria1990,Chubukov1994}. From Fig.~\ref{fig:SpinWaveCo}(b) we see that the in-plane transverse spin wave is along $\tilde x$ for all sub-lattices, which means that this Goldstone mode comes from fluctuations of $S^{\tilde x}$ at the $\Gamma$ point. The corresponding dynamical susceptibility is
   \beq
   \chi^{\tilde x\tilde x}(q,\Omega)=\frac{\rho_{\|}}{\Omega^2-v_{\|}^2q^2}.
   \eeq
The other two Goldstone modes correspond to out-of-plane spin wave fluctuations. Figs.~\ref{fig:SpinWaveCo}(c,d) show two independent modes. In one of them spins on the $A$ sublattice are fixed, and spins on the $B$ and $C$ sublattices rotate along $y$($\tilde y$) in the opposite direction. In the other mode, spins on the $B$ sublattice are fixed, and spins on the $A$ and $C$ sublattices rotate along $y$ in the opposite direction. The linear combinations of the two fluctuations yield two Goldstone modes with equal velocities around $\pm {\bf K}=(\pm 4\pi/3,0)$ in momentum space. The corresponding dynamical spin susceptibility is
  \beq
  \chi^{\tilde y\tilde y}(q\pm K,\Omega)=\frac{\rho_{\bot}}{\Omega^2-v_{\bot}^2q^2}.
  \eeq

In the rest of the paper, we drop the $\sim$ label for local coordinates unless there is ambiguity.

\begin{figure}[t]
\includegraphics[width=1\columnwidth]{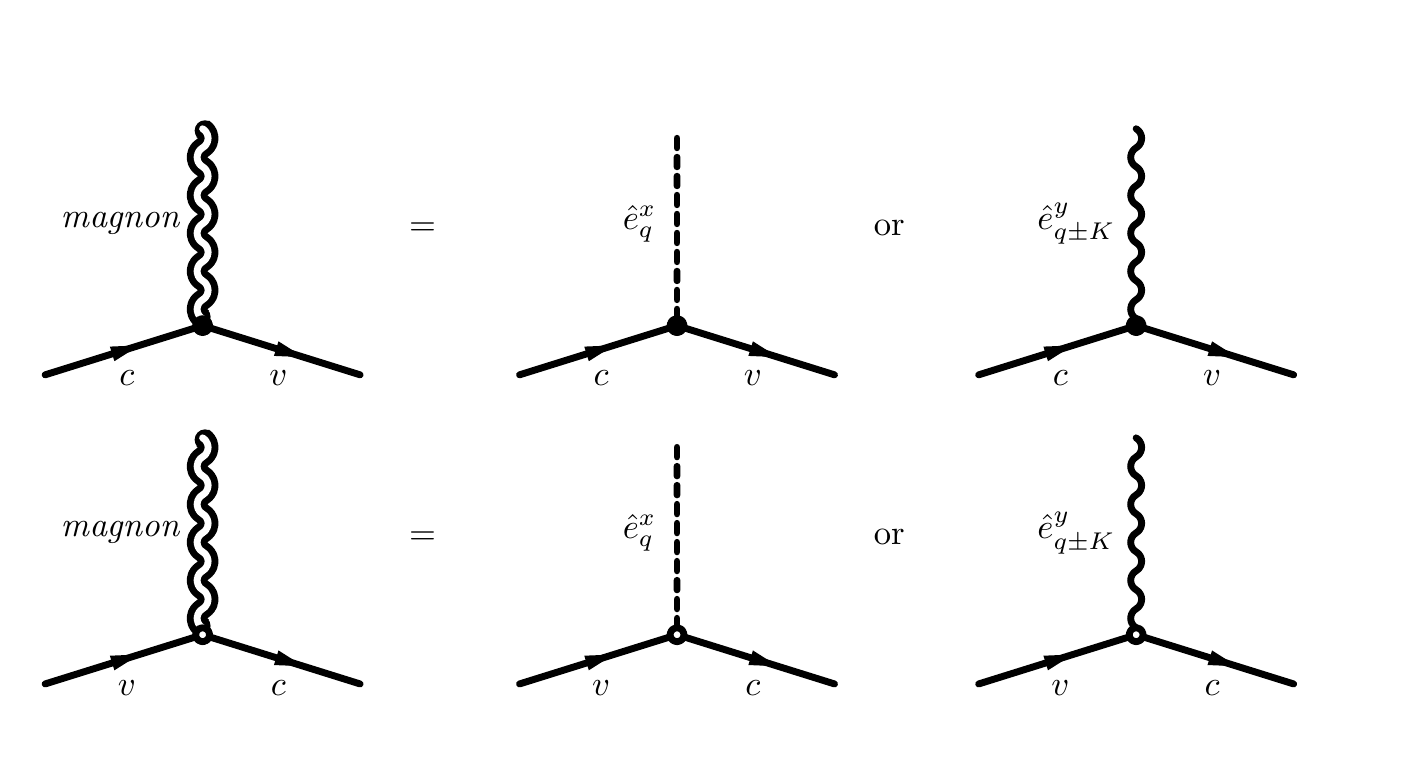}
\caption{ Magnon-fermion vertex. Double wavy line describes a magnon propagator with a generic momentum and spin component. 
Dashed and single wavy lines describe magnon propagators $\hat e^{x}_q$ near the $\Gamma$ point and for magnon propagator $\hat e^{y}_{q\pm K}$ near the $\pm K$ points, respectively. We use filled $\bullet$ (hollow $\circ$) circles to label vertices with incoming (outgoing) conduction fermion and outgoing (incoming) valence fermion.\label{fig:MagnonFermion}}
\end{figure}

In Hamiltonian approach,  static $\chi^{xx}(q, 0) = \chi^{xx}(q)$ and $\chi^{yy}(q, 0) = \chi^{yy}(q)$ determine the effective static interaction between fermions, mediated by magnons.
 To get this interaction, we take the Hubbard interaction
$\mathcal{H}_{int}=U\sum_i c^{\dagger}_{i,+} c_{i,+} c^{\dagger}_{i,-} c_{i,-}=\sum_i\frac{U}{2}(\hat n_{i,+}+\hat n_{i,-})-\frac{U}{2}\hat{S}_i^2$, dress it by  RPA renormalizations, and keep the spin part of the dressed interaction~\cite{Chubukov2014interaction} in the $\sigma^x \sigma^x$ and $\sigma^y \sigma^y$ channels. This yields
\begin{align}\label{eq:MagnonFermionInteraction0}
\mathcal{H}_{xx}&=-\frac{U^2}{N}\sum_{k,k',q}\chi^{xx}(q)c^{\dg}_{k+q}\sigma^x c_{k}\, c^{\dg}_{k'-q}\sigma^x c_{k'}\non\\
\mathcal{H}_{yy}&=-\frac{U^2}{N}\sum_{k,k',q}\chi^{yy}(q)c^{\dg}_{k+q\pm K}\sigma^y c_{k}\, c^{\dg}_{k'-q\mp K}\sigma^y c_{k'}
\end{align}
We now introduce the  magnon operators ($e^{x}_{q},\,e^{y}_{q\pm K}$) via
\begin{align}\label{eq:Magnon}
i\int \diff t e^{i \Omega t}\langle T e^{\tilde x}_{-q}(t)e^{\tilde x}_q(0)\rangle &=\chi^{\tilde x\tilde x}(q,\Omega)\non\\&=\frac{\rho_{\|}}{\Omega^2-v_{\|}^2q^2}\non\\
i\int \diff t e^{i \Omega t}\langle T e^{\tilde y}_{-q\mp K}(t)e^{\tilde y}_{q\pm K}(0)\rangle &=\chi^{\tilde y\tilde y}(q\pm K,\Omega)\non\\&=\frac{\rho_{\bot}}{\Omega^2-v_{\bot}^2q^2}
\end{align}
A little experimentation shows that Eq.~\eqref{eq:MagnonFermionInteraction0} are reproduced if we set magnon-fermion interaction to be
\begin{align}
\mathcal{H}_{m-f}=-U\sqrt{\frac{2}{N}}\sum_{k,q}\Big(c^{\dg}_{k+q}\sigma^x c_{k}\, \hat e^{x}_q + c^{\dg}_{k+q\pm K}\sigma^y c_{k}\, \hat e^{y}_{q\pm K}\Big)
\end{align}
In terms of the conduction and valence fermions, the interaction near the hot spot can be approximated as
\begin{align}\label{eq:MagnonFermionInteraction}
&\mathcal{H}_{m-f}=-U\sqrt{\frac{2}{N}}\sum_{k,q}\, \hat e^{x}_q (\gamma^{c\,\dg}_{k+q} \gamma^v_{k}+\gamma^{v\,\dg}_{k+q} \gamma^c_{k}) \non\\
&  -iU\sqrt{\frac{2}{N}}\sum_{k,q}\, \hat e^{y}_{q\pm K} (\gamma^{c\,\dg}_{k+q\pm K} \gamma^v_{k}-\gamma^{v\,\dg}_{k+q\pm K} \gamma^c_{k}).
\end{align}
We present this interaction graphically in Fig.~\ref{fig:MagnonFermion}. We use a double wavy line ($\aquarius$) for magnon propagator and use a filled circle ($\bullet$) for magnon-fermion vertex with outgoing valence fermion ($\gamma^{v\dg}$) and incoming conduction fermion ($\gamma^{c}$), and  an empty circle ($\circ$)
for the vertex with outgoing conduction fermion ($\gamma^{c\dg}$) and incoming valence fermion ($\gamma^{v}$).  From Eq.~\eqref{eq:MagnonFermionInteraction}, the magnon-fermion vertex for $\hat e^y$ is purely maginary, and is of  opposite sign for $\bullet$ and $\circ$ vertices. This turns out important when we calculate the full Green's function at the two-loop and  higher orders.
The interaction terms involving only conduction or only valence fermions are small in $q$ and will not be relevant to our analysis. The presence of $q$ in these terms is consistent with the Adler principle, which states that the interaction between Goldstone bosons and fermions from the same branch should be of gradient type, to preserve the Goldstone theorem (see Ref.~\cite{Chubukov1997} for more discussions). Note that the strength of magnon-fermion interaction is of order Hubbard $U$.

%========================================================
%========================================================
\section{The full fermionic Green's function in the SDW state}
\label{sec:3}
We now use the expressions for the quadratic SDW Hamiltonian, Eq.~\eqref{eq:Hmf}, and the magnon-fermion interaction, Eq.~\eqref{eq:MagnonFermionInteraction}, and obtain the expression for the full fermionic propagator $G^{c,v}(k_{hs},\omega)$ at a finite temperature $T$.  We show explicitly that the leading corrections come from the exchange of thermal transverse spin wave fluctuations.  These corrections are logarithmically singular in quasi-2D systems, and n-loop correction scales as $|\log\epsilon|^n$, where, we remind,  $\epsilon$ measures the deviation from pure two-dimensionality and serves as the infrared cutoff to regularize the divergence.

The fully renormalized Green's function at one of the hot spots is expressed as
\begin{widetext}
\begin{align}\label{eq:FullGreen}
G^{c,v}(k_{hs},i\omega_n,z)=G^{c,v\,(0)}(k_{hs},i\omega_n)\sum_{n=0}^{\infty}\mathcal{C}_n(z) (\beta U^2)^n
\big [G^{v,c\,(0)}(k_{hs},i\omega_n)G^{c,v\,(0)}(k_{hs},i\omega_n)\big ]^{n},
\end{align}
\end{widetext}
where the combinatoric factor $\mathcal{C}_n(z)$ increases factorially with $n$ and depends on the ratio of in-plane and out-of-plane spin-wave susceptibilities. The factor $\beta=\frac{\pi T}{\mathcal{A}_{BZ}}|(\chi_{\|}-2\chi_{\bot})||\log \epsilon|$ measures the strength of thermal fluctuations (see below).

To simplify the presentation below, we express the fully renormalized Green's function as
\begin{widetext}
\beq
G^{c,v}(k_{hs},i\omega_n,z)=G^{c,v\,(0)}(k_{hs},i\omega_n)+[G^{c,v\,(0)}(k_{hs},i\omega_n)]^2\tilde \Sigma(k_{hs},i\omega_n),
\eeq
\end{widetext}
where ${\tilde \Sigma}(k_{hs},i\omega_n)$ can be evaluated order by order in terms of $\beta$. We will use ${\tilde \Sigma}^{(m)}(k_{hs},i\omega_n)$ to label the $m^{th}$-loop correction. 
Note that ${\tilde \Sigma}^{(m)}(k_{hs},i\omega_n)$ is not equivalent to the $m^{th}$-loop self-energy as it includes both irreducible and reducible diagrams, which we will count on equal footings. Below we still use the term ``self-energy", but in reference to $\tilde \Sigma$.

\subsection{Perturbation theory at one-loop order}\label{sec:OneLoop}

\begin{figure}[bth!]
\includegraphics[width=1\columnwidth]{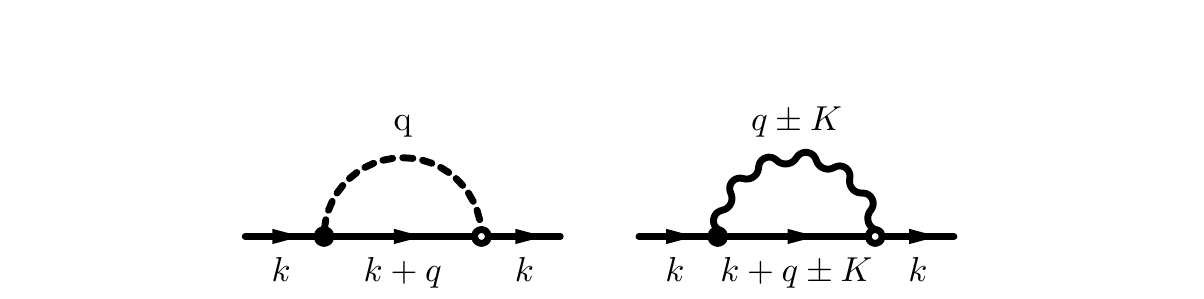}
\caption{One-loop self-energy diagrams from the exchange of thermal transverse spin fluctuations.} \label{fig:OneLoop}
\end{figure}
The fermion self-energy in Matsubara frequency at one-loop order is
\begin{align}
&\tilde \Sigma^{c,v\, (1)}(k,i\omega_n)=\non\\
&- U^2 \frac{T}{N}\sum_{q,m,j}\chi^{jj}(q,i\Omega_m)G^{v,c\,(0)}(k+q,i\Omega_m+i\omega_n)
\end{align}
The two singular contributions to $\Sigma^{c,v\, (1)}(k,i\omega_n)$ come from $xx$ and $yy$ components of the susceptibility. The contribution from $\chi^{xx}(q,i\Omega)$ to the leading logarithmical order is
\begin{align}\label{eq:OneLoop}
&\tilde\Sigma^{c,v\, (1a)}(k,i\omega_n)\non\\
&=-U^2 \frac{T}{N}\sum_{q,m}\frac{\rho_{\|}}{(i\Omega_m)^2-v_{\|}^2q^2}\frac{1}{i\Omega_m+i\omega_n-E^{v,c}_{k+q}}\non\\
&\approx -U^2 \frac{T}{N}\sum_{q,m=0}\frac{\rho_{\|}}{-v_{\|}^2q^2}\frac{1}{i\omega_n-E^{v,c}_{k+q}}\non\\
&=U^2\frac{\rho_{\|} T}{v_{\|}^2}\int\frac{\diff^2 q}{\mathcal{A}_{BZ}} \frac{1}{q^2}\frac{1}{i\omega_n-E^{v,c}_{k+q}}\non\\
&\approx U^2\frac{\pi\rho_{\|} T}{v_{\|}^2\mathcal{A}_{BZ}}|\log \epsilon |\frac{1}{i\omega_n-E^{v,c}_{k}}= \frac{\beta_1 U^2}{i\omega_n-E^{v,c}_{k}},
\end{align}
where we define $\beta_1=\frac{\pi\rho_{\|} T}{v_{\|}^2\mathcal{A}_{BZ}}|\log \epsilon|=\frac{\pi\chi_{\|} T}{\mathcal{A}_{BZ}}|\log \epsilon|$.
The contributions from non-zero bosonic Matsubara frequencies are finite, and we neglect them. The contribution from  $\chi^{yy}(q\pm K,i\Omega)$ is, similarly,
\begin{align}\label{eq:Oneloopb}
\tilde\Sigma^{c,v\, (1b)}(k,i\omega_n)&\approx U^2\frac{\pi\rho_{\bot} T}{v_{\bot}^2\mathcal{A}_{BZ}}|\log \epsilon| \frac{1}{i\omega_n-E^{v,c}_{k\pm K}}\non\\
&=\frac{\beta_2 U^2}{i\omega_n-E^{v,c}_{k\pm K}},
\end{align}
where $\beta_2=\frac{\pi\rho_{\bot}T}{v_{\bot}^2\mathcal{A}_{BZ}}|\log \epsilon|=\frac{\pi\chi_{\bot} T}{\mathcal{A}_{BZ}}|\log \epsilon|$.
At a hot spot, ${\bf k} = {\bf k}_{hs}$, $E^{v,c}_{{\bf k}_{hs}} = E^{v,c}_{{\bf k}_{hs} + {\bf K}} = \Delta$.
The sum of the two singular contributions then gives $\tilde\Sigma^{c,v\, (1)}(k,i\omega_n)=\frac{(\beta_1+2\beta_2) U^2}{i\omega_n-E^{v,c}_{k_{hs}}}$. We will see later that it is convenient to define $\beta=|\beta_1-2\beta_2|$ and  re-express $\tilde\Sigma^{c,v\, (1)}(k,i\omega_n)$ as
$\tilde\Sigma^{c,v\, (1)}(k,i\omega_n) = \Big(\frac{\beta_1+2\beta_2}{\beta}\Big)\beta U^2 G^{v,c\,(0)}(k_{hs},i\omega_n)$.

%========================================================
\subsection{Perturbation theory at two-loop order}
\begin{figure}[bth!]
\centering
\subfigure[]{\includegraphics[width=1\columnwidth]{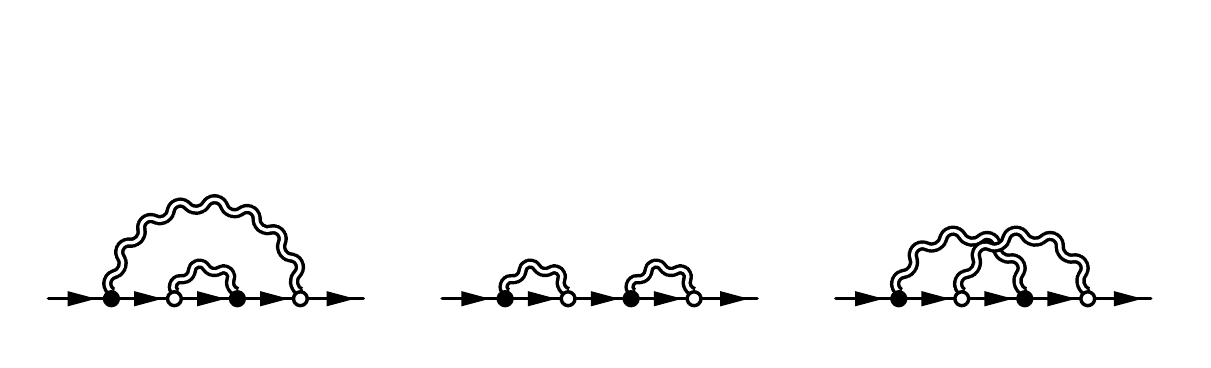}}
\subfigure[]{\includegraphics[width=0.9\columnwidth]{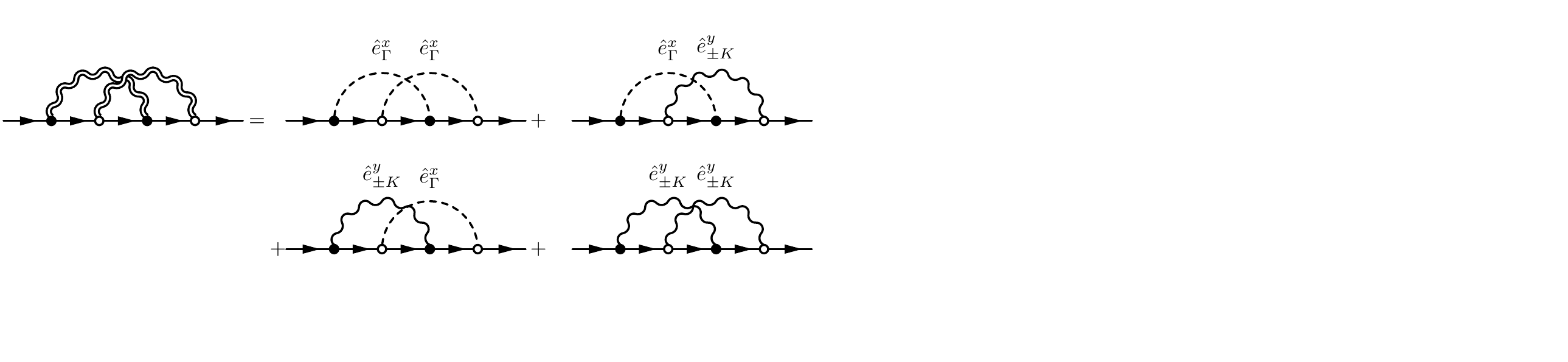}}
\caption{(a) The generic structure of two-loop diagrams. (b) The two-loop crossing diagrams from three magnon Goldstone modes. The overall factors in these diagrams are, from left to right and top to bottom, $\beta_1^2$, $-2\beta_1\beta_2$, $-2\beta_1\beta_2$, $(-2\beta_2)^2$.} 
\label{fig:TwoLoop}
\end{figure}

The self-energy at the two-loop order is obtained by summing over  $27=3^2 \, 3!!$ diagrams, where $(2n-1)!!\rightarrow 3!!$ counts the number of diagrams with different topology [see Fig.~\ref{fig:TwoLoop}(a)]. As there are three Goldstone modes, there are  $3^n\rightarrow 3^2$ diagrams in each topology. The overall factor for each diagram is  $\beta_{1}$ or $\beta_2$, like for one-loop diagrams, but the sign is either plus or minus. To explain the origin of sign alternation, consider the crossing diagram in Fig.~\ref{fig:TwoLoop}(b) as an example.
Because the magnon-fermion vertex for $\hat e^{y}$ is imaginary [see Eq.~\eqref{eq:MagnonFermionInteraction}], one can show that the two-loop Green's function from magnon propagator of $\hat e^{y}$ (wavy line) has prefactor $1=(\pm i )(\mp i )$ when ending with vertices of the opposite type ($\circ\aquarius\bullet$ or $\bullet\aquarius\circ$) as the magnon-fermion vertices associated with $\hat e^{y}$ contribute to a term $(i \gamma^{c \dg}\gamma^v)(-i \gamma^{v \dg}\gamma^c)$ in the expansion. Whereas it has prefactor $-1=(\pm i )(\pm i )$ when ending with vertices of the same type ($\bullet\aquarius\bullet$ or $\circ\aquarius\circ$) as the term takes a form $(i \gamma^{c \dg}\gamma^v)(i \gamma^{c \dg}\gamma^v)$ or $(-i \gamma^{v \dg}\gamma^c)(-i \gamma^{v \dg}\gamma^c)$. On the other hand, magnon propagator of $\hat e^{x}$ (dashed line) has the same prefactor $U^2$ for both the two ways that vertices are connected. Following these  rules, we find that the two-loop self-energy from crossing diagrams is
\begin{align}
&\tilde\Sigma^{c,v\, (2,crossing)}(k,i\omega_n)=(\beta_1-2\beta_2)^2\times\non\\
&\quad(U^2)^2 G^{v,c\,(0)}(k_{hs},i\omega_n)^2G^{c,v\,(0)}(k_{hs},i\omega_n).
\end{align}
Similarly, we find that the self-energy from non-crossing diagrams [the last two diagrams in Fig.~\ref{fig:TwoLoop}(a)] is
\begin{align}
&\tilde\Sigma^{c,v\, (2,non-crossing)}(k,i\omega_n)=2(\beta_1+2\beta_2)^2\times\non\\
&\quad (U^2)^2 G^{v,c\,(0)}(k_{hs},i\omega_n)^2G^{c,v\,(0)}(k_{hs},i\omega_n).
\end{align}
The total self-energy at the two-loop order
is
\begin{align}
&\tilde\Sigma^{c,v\, (2)}(k,i\omega_n)=\non\\
&\quad \mathcal{C}_2(\beta U^2)^2 G^{v,c\,(0)}(k_{hs},i\omega_n)^2G^{c,v\,(0)}(k_{hs},i\omega_n),
\end{align}
where $\beta=|\beta_1-2\beta_2|$ and  $\mathcal{C}_{2}=1+2\Big(\frac{\beta_1+2\beta_2}{\beta}\Big)^2$.
%========================================================
\subsection{perturbation theory at $n^{th}$-loop }\label{sec:nLoopGeneric}
Following the same procedure of computing the prefactors for the diagrams in which magnon propagators are connected by vertices of the same type or of opposite types, we find the self-energy at order $n$ to be
\begin{align}\label{eq:SelfEnergy1}
&\tilde\Sigma^{c,v\, (n)}(k,i\omega_n)=\non\\
&\quad\mathcal{C}_n(z)(\beta U^2)^n G^{v,c\,(0)}(k_{hs},i\omega_n)^nG^{c,v\,(0)}(k_{hs},i\omega_n)^{n-1},
\end{align}
where
\beq\label{eq:DefineZ}
z=\frac{\beta_1+2\beta_2}{\beta} = \frac{\chi_{\|} + 2\chi_{\bot}}{|\chi_{\|}-2\chi_{\bot}|},
\eeq
and the coefficient $\mathcal{C}_n(z)$ for $n=2 m$ even and $n=2m+1$ odd is expressed as (see Appendix~\ref{app:Fn} for details):
\begin{align}\label{eq:Fexpression}
\mathcal{C}_{2m}(z)&=\sum_{l=0}^{m}\big [\frac{(2m)\,!}{(2m-2l)\,!!}\big]^2\frac{z^{2l}}{(2l)\,!},\non\\
\mathcal{C}_{2m+1}(z)&=\sum_{l=0}^{m}\big [\frac{(2m+1)\,!}{(2m-2l)\,!!}\big]^2\frac{z^{2l+1}}{(2l+1)\,!},
\end{align}
where $l$ counts the number of magnon propagators ($2l$ for $n$ even, $2l+1$ for $n$ odd) which connect vertices of opposite type $\circ\aquarius\bullet$ or $\bullet\aquarius\circ$.

Summing up contributions from all loop orders, we find that the full fermionic Green's function can be expressed as
\begin{widetext}
\begin{align}\label{eq:FullGreen}
G^{c,v}(k_{hs},i\omega_n,z)=G^{c,v\,(0)}(k_{hs},i\omega_n)\sum_{n=0}^{\infty}\mathcal{C}_n(z)(\beta U^2)^n \big [G^{v,c\,(0)}(k_{hs},i\omega_n)G^{c,v\,(0)}(k_{hs},i\omega_n)\big ]^{n}.
\end{align}
\end{widetext}
where $\mathcal{C}_0=1$ (this also follows  from Eq.~\eqref{eq:Fexpression}, if we set $m=0$).

We see from Eq.~\eqref{eq:DefineZ} that the value of $z$ depends on microscopic details, which set the ratio of the susceptibilities. In the large $U$ limit, the Hubbard model is well approximated by the nearest-neighbor Heisenberg model~\cite{ChubukovMusaelian1994,*ChubukovMusaelian1995}. For the latter, spin-wave calculations done at large $S$ (the case reproduced by taking $2S$ flavors of fermions) yield $\chi_{\|}=\frac{2}{9\sqrt{3} J a^2}(1-\frac{0.449}{2S})$, $\chi_{\bot}=\frac{2}{9\sqrt{3} J a^2}(1-\frac{0.285}{2S})$~\cite{Chubukov1994}. Using these expressions we find $z=3-\frac{0.32}{S}+\mathcal{O}(1/S^2)$. At smaller $U$, the value of $z$ changes, because there appear additional terms in the effective spin Hamiltonian, and, in principle, can be any number. In particular, $z=1$ when $\chi_{\|}=0$ or $\chi_{\bot}=0$; $z=\infty$ when $\chi_{\bot}=\chi_{\|}/2$.

\begin{figure}%[bth!]
\centering
\subfigure[]{\includegraphics[width=0.9\columnwidth]{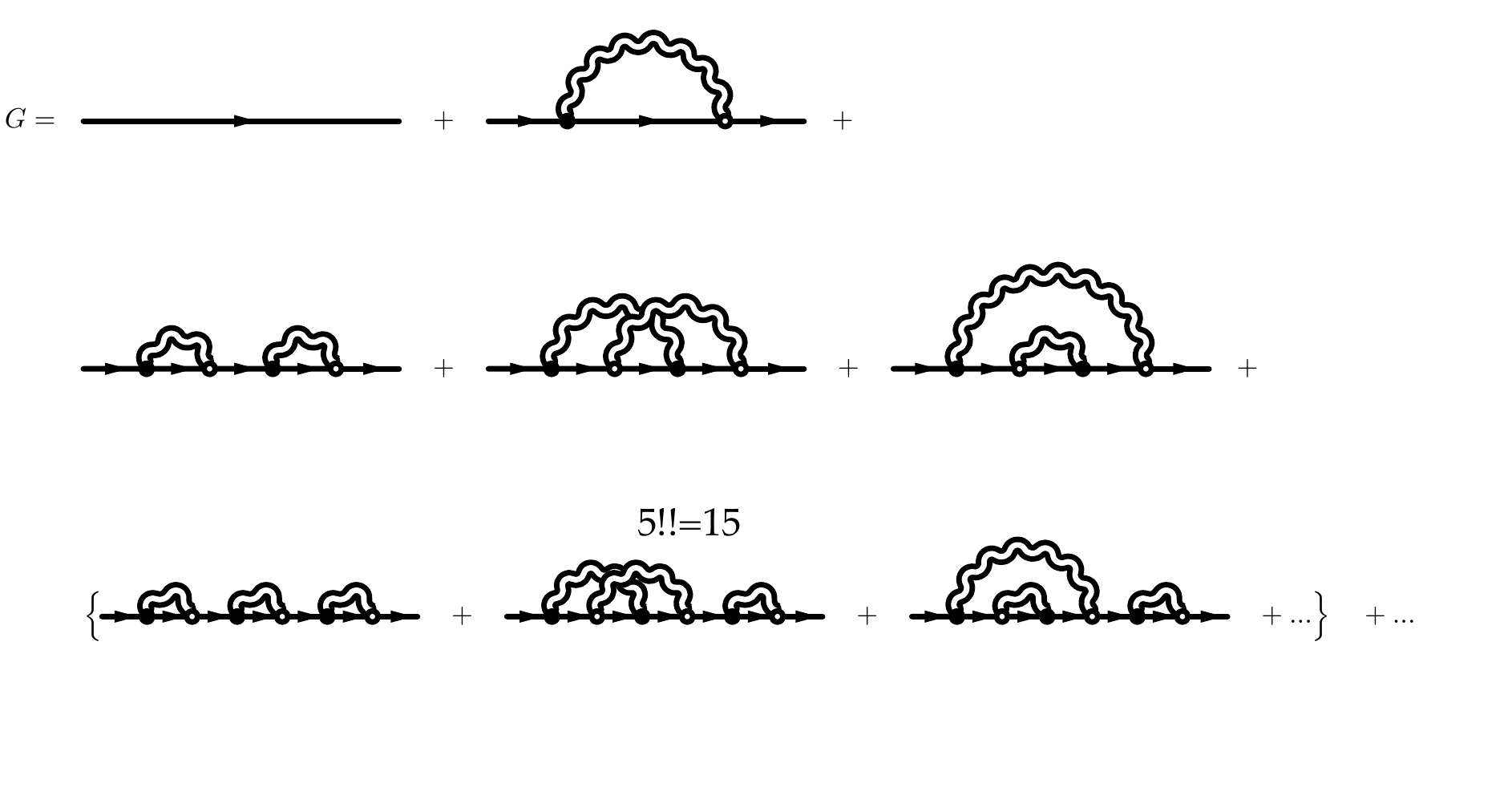}}
\subfigure[]{\includegraphics[width=0.9\columnwidth]{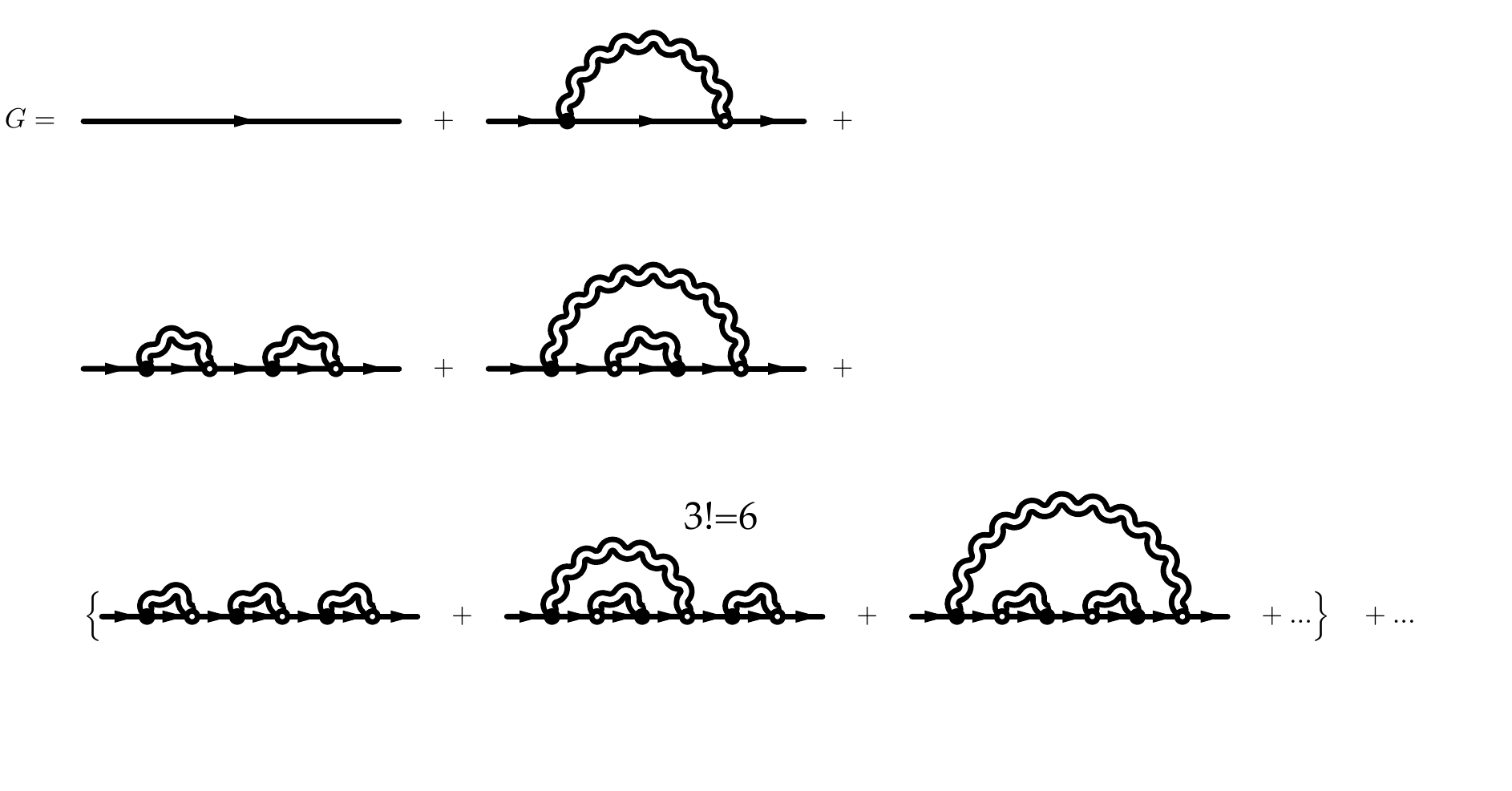}}
\caption{ The structure of the diagrammatic series for the cases of (a) $z=1$ and (b) $z=\infty$. 
\label{fig:zDiag}}
\end{figure}

For these two limiting cases, the combinatoric factor $\mathcal{C}_n(z)$ can be obtained in a closed form, as a function of $n$ (as opposed to the sum for a generic $z$, as in Eq. (\ref{eq:Fexpression})). For $z=1$, there is no distinction between magnon propagators which connect vertices of opposite types, $\circ\aquarius\bullet$, $\bullet\aquarius\circ$ or of the same type, $\circ\aquarius\circ$, $\bullet\aquarius\bullet$. For $n$-loop diagrams, there are $2n$ vertices, thus there are  $(2n-1)!!$ topologically distinct diagrams [see Fig.~\ref{fig:zDiag}(a)]. In this situation, $\mathcal{C}_n(z=1)=(2n-1)!!$. In the opposite limit $z \to \infty$, we need to introduce $\beta'$ via  $\beta=\frac{\beta'}{z}\rightarrow 0$, and keep $\beta'$ as a constant. The most relevant term in Eq.~\eqref{eq:Fexpression} at $z \to \infty$ is the one with $l=m$. The corresponding diagrams contain only magnon propagators connecting vertices of the opposite type ( $\circ\aquarius\bullet$ and  $\bullet\aquarius\circ$). At $n^{th}$-loop order, there are $n!$ topologically distinct diagrams [see Fig.~\ref{fig:zDiag}(b)], so $\beta^n \mathcal{C}_n(z=\infty)=n!\beta'^n$, i.e., $\mathcal{C}_n(z=\infty) \to n!$.

We note in passing that the structure of multi-loop corrections to a fermionic propagator for $z=1$ and for $z=\infty$ is the same as quasi-1D models with CDW order/fluctuations. The case $z=1$ is realized at half-filling, when the  ordering wave vector is  $Q=\pi$). The case $z=\infty$ is realized in generic filling~\cite{Sadovskii1974,*Sadovskii1974b,*Sadovskii1979,*[{For a detailed discussion of the formalism, see }]SadovskiiBook}. To our knowledge, there have been no prior analysis of a generic $z$, only specific cases have been considered. In our case, the value of $z$ is determined by ratio of $\chi_{\|}/\chi_{\bot}$ and can be arbitrary in the interval $[1,\infty)$. In the next section we analyze how the spectral function behaves for different $z$.
%========================================================
%========================================================
\section{The spectral function}\label{sec:4}
%========================================================
\subsection{Evaluation}\label{sec:4-1}
The spectral function is defined as $A^{c,v}(\khs,\omega)=-\frac{1}{\pi}\im G^{c,v}(k_{hs},\omega+i\delta)$, where $\omega$ is a real frequency.
Our goal is to evaluate $A^{c,v}(\khs,\omega)$  analytically,  starting from Eqs.~\eqref{eq:Fexpression} and \eqref{eq:FullGreen}.
The key technical challenge is to perform the summation over $n$ in Eq.~\eqref{eq:FullGreen} in a situation when $\mathcal{C}_n(z)$ is not expressed in a closed form. We note that
  because $\mathcal{C}_{n}(z)\sim \mathcal{O}(n!)$, a numerical computation of $A^{c,v}(\khs,\omega)$ is quite challenging on its own.

Our strategy  is to first sum over $l$ in Eq.~\eqref{eq:Fexpression}, and express $\mathcal{C}_{n}(z)$ in an integral form as
\begin{align}\label{eq:FexpressionMain}
\mathcal{C}_{n}(z)=\frac{n!}{2^{n}2\pi i}\oint^{(0+)}\diff v\, \big(\frac{1}{v}\big)^{n+1}\frac{(1+2zv)^{n}}{\sqrt{1-4v^2}}
\end{align}
for  $n\in \mathbb Z$ ($\oint^{(0+)}$ means the contour integral goes around the pole at $v=0$ counter-clockwisely, see Fig.~\ref{fig-11}(a)). Eq.~\eqref{eq:FexpressionMain} makes the analytic summation over $n$ in Eq.~\eqref{eq:FullGreen} possible, as $n$ only appears as an overall factor $n!$ and as an exponent in the integrand. Summing over $n$ and converting from Matsubara to real frequency ($i\omega_m \to \omega + i \delta$),
 we find the spectral function in the form of a single integral (see  Appendix~\ref{app:SpectralFunction} for details).
 \begin{widetext}
\begin{align}\label{eq:resultMain}
A^{c,v}(\khs,\omega)=\frac{1}{\pi}\Big|\frac{1}{\omega\mp \Delta}\Big|\int_{\frac{1}{(z+1)u_\omega}}^{\frac{1}{(z-1) \,u_\omega}}\diff t\, e^{-t}\, \frac{1}{\sqrt{(u_\omega\,t)^2-(1-u_\omega\,t\,z)^2}}\Theta(u_\omega),
\end{align}
\end{widetext}
where  $u_\omega=\frac{\beta U^2}{\omega^2-\Delta^2}$ and $\Theta(u_\omega)$ is the Heaviside step function.
Eq.~\eqref{eq:resultMain} is the main result of this paper.
In the next section  we analyze qualitative features of the spectral function for different $z$.

%========================================================
\subsection{Results}\label{sec:4-2}
We can extract from Eq.~\eqref{eq:resultMain} a few generic properties of the spectral function.
\begin{itemize}
\item The presence of the Heaviside step function $\Theta(\frac{\beta U^2}{\omega^2-\Delta^2})$ on the r.h.s.\ of Eq.~\eqref{eq:resultMain} means that $A^{c,v}(\khs,\omega)$ vanishes for $\omega\in (-\Delta,\Delta)$, i.e., the SDW order parameter defines the real gap at a hot spot.
\item At the point where SDW order disappers, $\Delta=0$ and $u_\omega=\frac{\beta U^2}{\omega^2}$. One can understand whether at this point the system displays a pseudogap behavior or a conventional Fermi liquid behavior by analyzing how the spectral function behaves at $\omega\sim0$. When $z>1$, the integral is bounded in the ultra-violet, and a straightforward analysis shows that $A^{c,v}(\khs,\omega)\sim \omega$, i.e., the maximum of $A^{c,v}(\khs,\omega)$ is at a finite frequency. This implies that the system displays a pseudogap behavior. At $z=1$, the upper bound of the integral becomes infinite ($\frac{1}{(z-1)u_{\omega}}\rightarrow\infty$). Then  $A^{c,v}(\khs,\omega=0)$ doesn't vanish. One can expand in $\omega$ and check that $A^{c,v}(\khs,\omega)$ has a maximum at $\omega=0$.  This is the expected behavior in an ordinary Fermi liquid. We verified the dichotomy between the cases $z >1$ and $z =1$ by analytical calculations for $z=1$ and $z=\infty$ and numerical calculations for an arbitrary $z\in (1,\infty)$, as we show below.
\end{itemize}

\begin{figure}%[H]%[bth!]
\centering
\subfigure[]{\includegraphics[width=0.75\columnwidth]{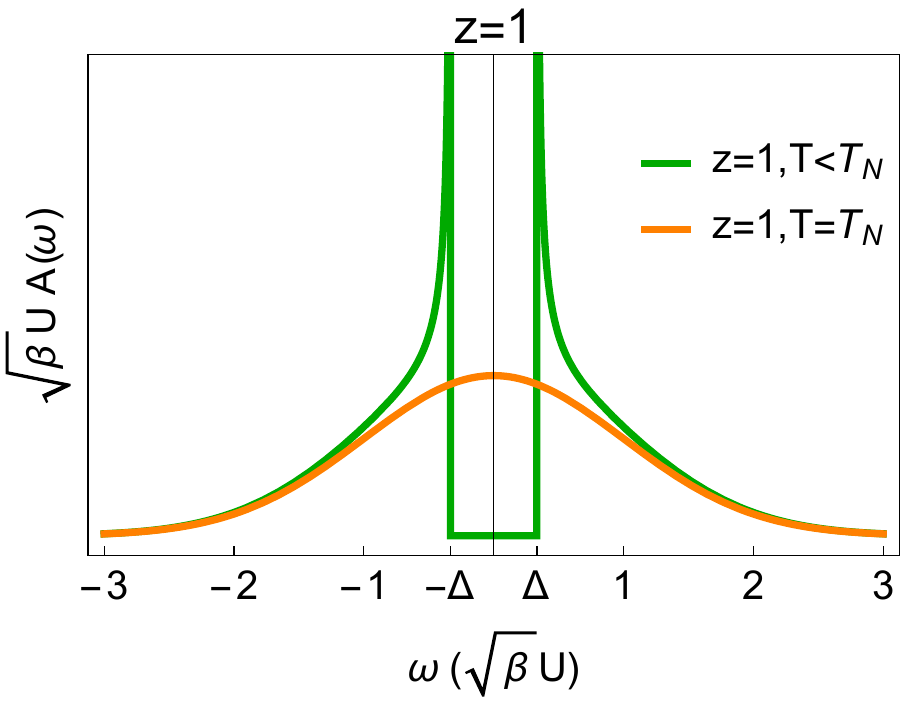}}
\subfigure[]{\includegraphics[width=0.75\columnwidth]{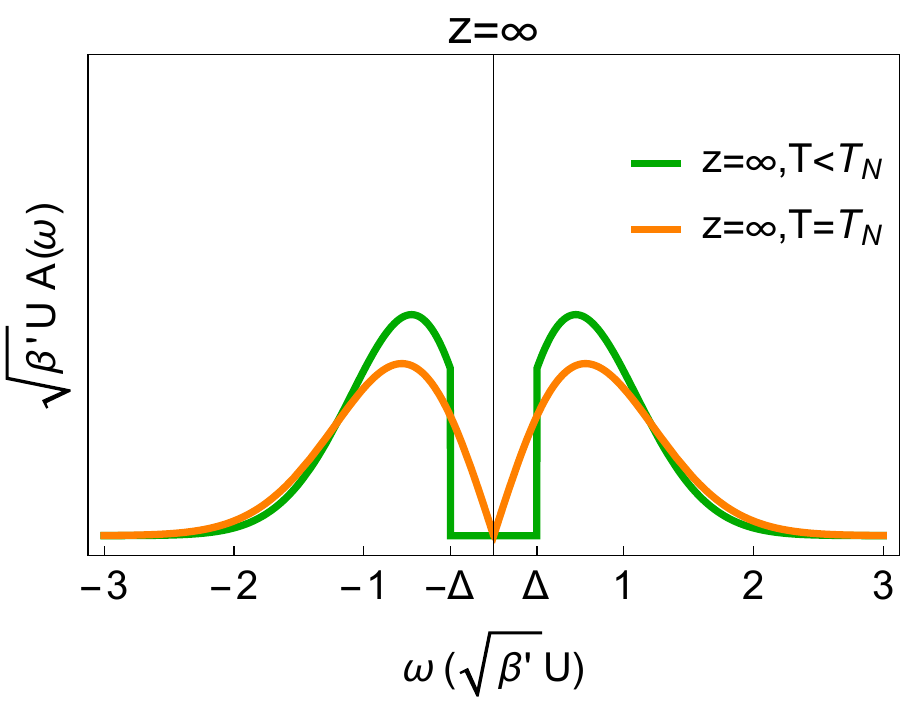}}
\subfigure[]{\includegraphics[width=0.75\columnwidth]{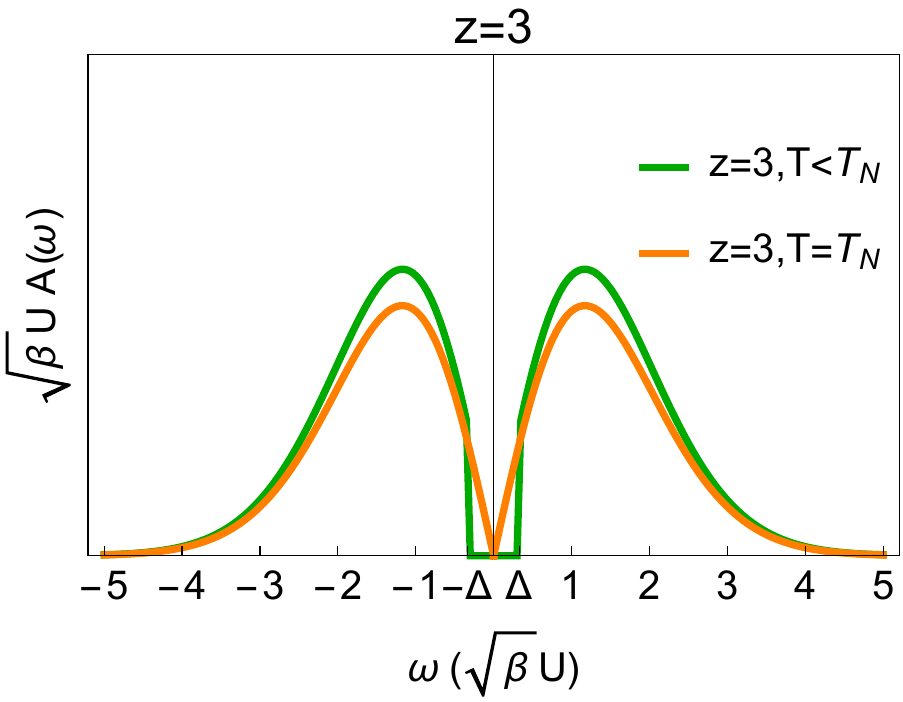}}
\caption{The spectral function at a hot spot for different $z=|\frac{\chi_{\|}-2\chi_{\bot}}{\chi_{\|}+2\chi_{\bot}}|$.  This spectral function includes the effects of series of scattering by transverse thermal fluctuations.
Green lines -- deep in the ordered state, $T \ll T_N$; orange lines --  at $T = T_N$, when $\Delta=0^{+}$.
Panels (a)-(c) are for $z=1$, $z = \infty$, and $z=3$.\label{fig:Spectralz}}
\end{figure}

\begin{figure}[H]%[bth!]
\centering
\includegraphics[width=0.9\columnwidth]{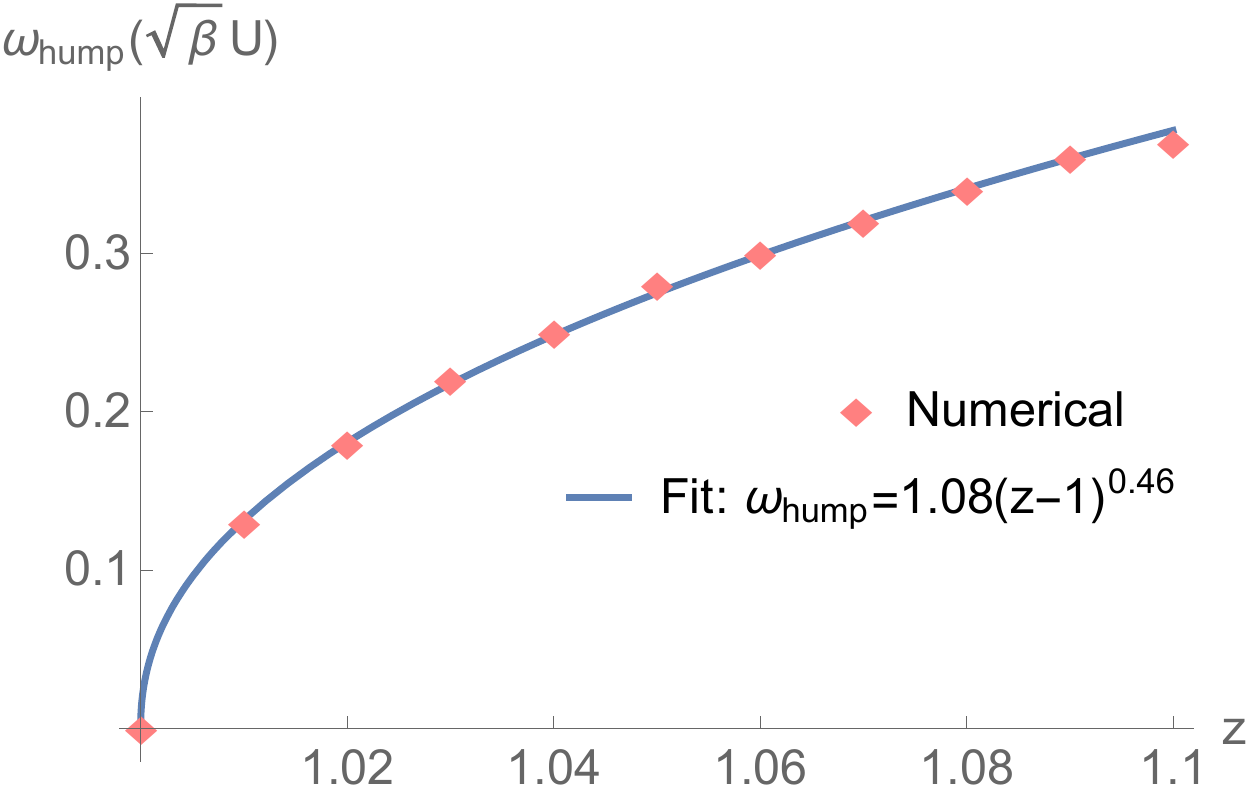}
\caption{The position of the hump, $\omega_{hump}$, as a function of $z$.} \label{fig-6}
\end{figure}

\emph{Analytical result at $z=1$.}~~ At $z=1$, the upper bound of the integral goes to $\infty$ and the integral in Eq.~\eqref{eq:resultMain}
can be evaluated analytically. The result is
\begin{align}
&A^{c,v}(\khs,\omega)\non\\
&=\frac{1}{\pi}\Big|\frac{1}{\omega\mp \Delta}\Big|\int_{\frac{1}{2u_{\omega}}}^{\infty}\diff t\, e^{-t}\, \frac{1}{\sqrt{2u_{\omega}t-1}}\Theta(u_{\omega})\non\\
&=\frac{1}{\pi}\Big|\frac{1}{\omega\mp \Delta}\Big|\int_{0}^{\infty}\diff \eta\, \frac{1}{2u_{\omega}}e^{-\frac{\eta+1}{2u_{\omega}}}\, \frac{1}{\sqrt{\eta}}\Theta(u_{\omega})\non\\
&=\frac{1}{\pi}\Big|\frac{1}{\omega\mp \Delta}\Big|\sqrt{\frac{\pi}{2u_{\omega}}}e^{-\frac{1}{2u_{\omega}}}\Theta(u_{\omega})\non\\
&=\Theta(|\omega|-\Delta)\sqrt{\frac{1}{2\pi\beta U^2}}\sqrt{\Big|\frac{\omega\pm \Delta}{\omega\mp \Delta}\Big|}e^{-\frac{\omega^2-\Delta^2}{2\beta U^2}},
\end{align}
The same result was obtained in Ref.~\cite{Sedrakyan2010}. We plot $A^{c,v}(\khs,\omega)$ in Fig.~\ref{fig:Spectralz}(a). Taking the limit $\Delta \to 0$, we obtain
 \beq
 A^{c,v}(\khs,\omega)=\sqrt{\frac{1}{2\pi\beta U^2}}e^{-\frac{\omega^2}{2\beta U^2}}.
 \eeq
We see that at $\Delta \to 0$ (i.e., at $T \to T_N$) the spectral function is peaked at $\omega=0$. This implies, as we anticipated,  that for $z=1$  thermal fluctuations  do not give rise to  SDW precursors. We emphasize that this could not be anticipated from the few first terms in loop expansion of the self-energy as these terms show little difference between $z=1$ and larger values of $z$. In the SDW phase, the spectral weight is zero at $\omega\in (-\Delta,\Delta)$, has a peak at $\omega=\pm (\Delta + 0)$, and gradually decays at higher frequencies, i.e., it does not show peak/dip/hump structure. This is indeed consistent with the absence of a pseudogap at $T = T_N$.

\textit{Analytical result at $z=\infty$.}~~The integral in Eq.~\eqref{eq:resultMain} can be evaluated analytically also at $z \to \infty$. As we discussed before, in this limit one should introduce $\beta'$ via $\beta=\frac{\beta'}{z}$ and keep $\beta'$ finite. Expanding the upper and lower bounds of the integral in Eq.~\eqref{eq:resultMain} as $\frac{1}{(z\mp1)u_{\omega}}=\frac{\omega^2-\Delta^2}{\beta' U^2}(1\mp \frac{1}{z})$, and subsituting into Eq.~\eqref{eq:resultMain}, we obtain after some algebra
\begin{align}
A^{c,v}(\khs,\omega)=\Theta(|\omega|-\Delta)\frac{|\omega\pm \Delta|}{\beta' U^2}e^{-\frac{\omega^2-\Delta^2}{\beta' U^2}}.
\end{align}
This result is also obtained if we  replace $\beta^n \mathcal{C}_n(z=\infty)$ by $n!\beta'^n$ and directly sum up over $n$. We show $A^{c,v}(\khs,\omega)$ in Fig.~\ref{fig:Spectralz} (b).

At $T = T_N$, we have
 \beq
 A^{c,v}(\khs,\omega)=\frac{\omega}{\beta' U^2}e^{-\frac{\omega^2}{\beta' U^2}}.
  \eeq
We see that now the spectral function scales as $\sim \omega$ at small frequencies and has a maximum (a hump) at a frequency $\omega \sim \sqrt{\beta'}U$. This implies that at $T= T_N$ the system retains memory about an SDW state. As $T $ increases above $T_N$,  the maximum in the spectral function remains at a finite frequency over some range of $T$. This is a canonical pseudogap (precursor to magnetism) behavior.

\textit{Numerical results at $z\in (1,\infty)$.}~~We found numerically that for any $z >1$ the spectral function behaves qualitatively the same as at $z=\infty$. Fig.~\ref{fig:Spectralz} (c) shows the spectral function for $z=3$ (the value of $z$ for our system at large $U$ and large spin S, i.e., large number of fermionic flavors). As we said above, this ``universality" is expected because the upper bound of the integral in Eq.~\eqref{eq:resultMain} is finite, as long as $z>1$. Then Eq.~\eqref{eq:resultMain} can be simplified as
\begin{widetext}
\begin{align}\label{eq:Spectral2}
A^{c,v}(\khs,\omega)&=\frac{1}{\pi}\Big|\frac{1}{\omega\mp \Delta}\Big|\int_{\frac{1}{(z+1)u}}^{\frac{1}{(z-1) \,u}}\diff t\, e^{-t}\, \frac{1}{\sqrt{\big(\frac{1}{(z-1) \,u}-t\big)\big(t-\frac{1}{(z+1) \,u}\big)(z-1)u(1+z)u}}\Theta(u)\non\\
&=\Theta(|\omega|-\Delta)\frac{|\omega\pm \Delta|}{\pi\beta U^2\sqrt{(z-1)(z+1)}}\int_{\frac{1}{(z+1)u}}^{\frac{1}{(z-1) \,u}}\diff t\,\frac{ e^{-t}}{\sqrt{\big(\frac{1}{(z-1) \,u}-t\big)\big(t-\frac{1}{(z+1) \,u}\big)}}.
\end{align}
\end{widetext}
At $T=T_N$, $A^{c,v}(\khs,\omega)$ scales linearly with $\omega$ at small frequencies. This necessarily implies pseudogap behavior in some range of temperatures above $T_N$. We found numerically that a maximum (a hump) is located at $\omega\sim\sqrt{\beta}U$.

To understand the behavior near $z=1$ we calculated numerically the position of the hump ($\omega_{hump}$) at $T = T_N$, as a function of $z$. We show the result in Fig.~\ref{fig-6}. Observe that $\omega_{hump}$ increases quite rapidly, as $(z-1)^{0.46}$. This indicates that the pseudogap feature at $T=T_N$ is quite robust, as long as $z>1$, while $z=1$ should be viewed as a special case. However, we should note that varying the value of $z$ changes the temperature where the hump starts to show up in the spectral function. In this paper, though an accurate analysis that maps the temperature $T\sim \beta$ with the order parameter $\Delta$ was not done, we can see the trend from a qualitative argument. On the one hand, $\omega_{hump}\sim (z-1)^{0.46} \sqrt{\beta} U$ increases with temperature. On the other hand, $\Delta$ decreases with temperature. As the hump shows up when $\omega_{hump}\gtrsim \Delta$, a higher temperature is needed to compensate for the smallness of $(z-1)^{0.46}$ as $z\rightarrow 1$.

%========================================================
\subsection{Additional considerations}\label{sec:4-3}

First, we note that the spectral function, which we obtained,  doesn't have the coherent peak at $\omega=\pm \Delta$.  This is an artifact of the approximation, in which we only include logarithmically singular contributions to the self-energy from thermal spin fluctuations. A coherent peak at $\omega=\pm \Delta$ is recovered once we add contributions to the self-energy from quantum fluctuations and  non-singular self-energy piece from thermal fluctuations. This issue has been addressed in Ref. ~\cite{Sedrakyan2010}. We show the result of including these additional terms into the self-energy (and the spectral function) by  dashed line in Fig.~\ref{fig:summaryPG} (b).

 Second, in this paper we considered the self-energy due to exchange of transverse spin-wave fluctuations.  As we said, such an exchange gives rise to series of logarithmically singular self-energy terms. Deep inside the SDW phase, longitudinal spin fluctuations are gapped and contribute only little to the self-energy. However, as the temperature increases towards $T_N$, the gap in the longitudinal fluctuations gets reduced. A more careful study at $T \lesssim T_N$ should take into account the contribution from longitudinal channel. Such analysis has been performed for a system on a square lattice~\cite{Schmalian1998,*Schmalian1999,Sedrakyan2010}, and the conclusion was that longitudinal fluctuations enhance the tendency towards precursor behavior. In mathematical terms, this happens because the combinatoric factor changes from $n!$ (only transverse fluctuations, $z\to \infty$), to $(2n+1)!!$ (transverse and longitudinal fluctuations). We note in this regard that our $\mathcal{C}_n(z)$  does not become  $(2n+1)!!$ for any $z$. As a result, the behavior of spectral function near $\omega=0$ changes from $\sim \omega$ to $\sim \omega^2$, and the energy of the hump increases. In our case (fermions on a triangular lattice) the analysis of the self-energy from longitudinal fluctuations is more involved than in the case of square lattice, and we refrain from making a definite prediction.  Still, it is possible that longitudinal fluctuations induce some precursor behavior near $T_N$  even for $z=1$.

Third, it is interesting to compare our non-perturbative solution for the spectral function with a perturbative solution in which one restricts with the one-loop self-energy. In our computational approach, this implies that one includes irreducible diagrams for the Green's function at one-loop order and only reducible diagrams at higher orders. The perturbative result for the spectral function is then
\begin{align}
A^{pert}_{k}(\omega) &=\re G^{(0)}\times \im \sum_{n=0}^{\infty} u_{\omega}^n\non\\
&=\re G^{(0)} \im \frac{1}{1-u_{\omega}}=\pi\delta(1-u_{\omega})\re G^{(0)}
\end{align}
where, we remind, $u_{\omega}=\frac{\beta U^2}{\omega^2-\Delta^2}$. This result holds for any value of $z$.
We  see that $A^{pert}_{k}(\omega)$ has two $\delta$-functional peaks at $\omega=\pm \sqrt{\Delta^2+\beta U^2}$. The peak frequency remains finite at $\Delta =0$, which implies that some evidence for a precursor to magnetism appears already within the perturbation theory. 
However, the full expression (Eqs.~\eqref{eq:spectral} and~\eqref{eq:sum3} in Appendix~\ref{app:SpectralFunction}), is more involved:
\begin{align}
A_{k}(\omega) &=\re G^{(0)}\times \im \sum_{n=0}^{\infty}\mathcal{C}_n(z)u_{\omega}^{n}\non\\
&=\re G^{(0)}\im \int_{\frac{1}{(z+1)u_{\omega}}}^{\frac{1}{(z-1)u_{\omega}}}\diff t\, e^{-t}\, \frac{1}{\sqrt{1-4v_0^2}}\frac{1}{1-u_{\omega}\,t\,z}.
\label{aaaa}
\end{align}
The most essential difference is that the full $A_{k}(\omega)$ has contributions not only from the pole but also from the branch cut (the  $\frac{1}{\sqrt{1-4v_0^2}}$ term). The branch cut contribution gives rise to the incoherent part of the spectral function.  Combining the pole and the branch cut contributions, 
%Only with the last contribution we  
%\MY{Due to the combinatoric factor $\mathcal{C}_n(z)$, the non-analyticity comes from the branch cut in $\frac{1}{\sqrt{1-4v_0^2}}$ rather than the pole in $ \frac{1}{1-u_{\omega}}$. 
%Only then 
we obtain both the gap below $\Delta (T)$ and the hump at an energy $\Delta (0) \approx U/2$ (and we recall that with respect to the renormalized $\mu(T)$, the hump is at energy of order $J$).
%========================================================
%========================================================
\section{Summary}\label{sec:discussions}
In this paper we studied the effects of thermal fluctuations on the spectral function of hot fermions on a triangular lattice, in the $120^\circ$ SDW state  (the ordering momentum is ${ \bf K} =  (4\pi/3,0)$).

We argued that the exchange of static Goldstone bosons between fermions in the valence and the conduction band gives rise to logarithmically singular self-energy corrections. We obtained fully renormalized Green's function by summing up infinite series of thermal self-energy diagrams. In this sense, we went beyond perturbation theory.

The key goal of our study was to understand whether the exchange of static thermal bosonic fluctuations necessarily gives rise to pseudogap behavior, or the system may display a conventional Fermi liquid behavior despite that self-energy corrections are logarithmically singular. We argued that one can address this issue by studying fermions on a triangular lattice. Specifically, we  showed that the contributions from in-plane and out-of-plane spin-wave fluctuations are not equivalent, and the strength of the self-energy renormalizations depends on the ratio of in-plane and out-of-plane spin-wave susceptibilities $\chi_{\|}/\chi_{\bot}$.  This ratio is an input parameter for low-energy theory, and by varying it one can study the changes in the structure of diagrammatic series for the self-energy. When $\chi_{\|}/\chi_{\bot}\sim 1$, our calculations show that the behavior of the spectral function for a fermion at a hot spot (${\bf k}_{hs}$ and ${\bf k}_{hs} + {\bf K}$ are both on the Fermi surface) is similar to that for the case of a square lattice and collinear $(\pi,\pi)$ SDW order: there is a real gap below $\Delta(T)$ and a maximum (hump) at an energy of order Hubbard $U$ (see Fig.~\ref{fig:summaryPG}). The hump persists at $T= T_N$, where $\Delta$ vanishes, and survives in some range of $T$ above $T_N$. In this range the system displays a pseudogap behavior. On the other hand, when $\chi_{\|}/\chi_{\bot}\ll 1 $ or $\chi_{\|}/\chi_{\bot}\gg 1 $, the pseudogap behavior exists only near $T_N$. In the limiting case when $\chi_{\|}/\chi_{\bot}=0 $ or $\chi_{\|}/\chi_{\bot}=\infty$, there is no pseudogap behavior at any $T > T_N$, despite that perturbative self-energy corrections are logarithmically singular.

The calculations, which we presented in this paper, can be readily extended to other microscopic models, and  our results show that by looking at the structure of perturbation series one would be able to immediately conclude whether singular self-energy corrections lead to a pseudogap behavior,  or to an ordinary Fermi liquid behavior.

\section{Acknowledgements}
We thank L.\ Fu, A.\ Georges, E.\ Gul, E.\ Kozik, J.P.F.\ LeBlank, J.\ Schmalian, F.\ Simkovic and Z.\ Zhang for helpful discussions. The work was supported by the NSF DMR-1523036. M.Y.\ also acknowledges support from Louise Dosdall Fellowship from the University of Minnesota.
%========================================================
%========================================================
\onecolumngrid
\clearpage
\appendix
%========================================================
\section{Goldstone modes}\label{app:Goldstone}
In this appendix, we give more mathematical details of how the momentum and spin structure of the low energy bosonic collective modes in the SDW state, i.e., the magnon Goldstone modes, can be obtained by studying the eigenstates of its corresponding linear spin wave Hamiltonian. As we show in the main text, these universal properties of the Goldstone modes are enough to obtain the effective magnon-fermion interaction.

To be precise, as our starting point is the SDW state out of itinerant fermions, the collective modes in the interacting fermion system, magnons in this case, should be obtained in principle by calculating the spin correlation function. On the other hand, as the low energy magnons are essentially the Goldstone modes which are uniquely determined by the pattern of the spontaneous spin rotation symmetry breaking, the universal properties of the Goldstone modes can be obtained from other models that are adiabatically connected to the Hubbard model in the large $U$ limit, e.g.\ the isotropic Heisenberg model. For the square lattice Hubbard model, it has been checked that the spectrum calculated from the two methods match over the whole Brillouin~\cite{Schrieffer1989,ChubukovFrenkel1992}. For the triangular lattice Hubbard model, though it turns more tricky at higher energy~\cite{Tremblay1995}, the matching should work in principle for the low energy Goldstone modes for the reason we explained above. We also checked that it is indeed the case. In the following, we work with the nearest neighbor isotropic Heisenberg antiferromagnetic model using linear spin wave analysis. The large $S$ spin wave expansion can be reproduced by taking $2S$ fermion flavors in the Hubbard model.

The global and local coordinates are set up as shown in Fig.~\ref{fig:SpinWaveCo}(a). To obtain the linear spin wave Hamiltonian, we express the spin operators in terms of Holstein-Primakoff bosons $a,\,a^{\dagger}$. In the local coordinate where the magnetic order is along $\tilde z$, the spin operator is:
\begin{align}
S_{{r}}^+(\tilde {z}) =\sqrt{2S}\sqrt{1-\frac{a_{{r}}^{\dagger}a_r}{2S}}a_r,\quad
S_{{r}}^-(\tilde {z}) =\sqrt{2S}a_r^{\dagger}\sqrt{1-\frac{a_r^{\dagger}a_r}{2S}},\quad
S_{{r}}^z(\tilde {z}) =S-a_r^{\dagger}a_r
\label{appeq:HP}
\end{align}
The spin operators in the global coordinate can be expressed as:
 \begin{align}
 S_{i}^{x}=S^{x}(\tilde {z})\cos\theta_i+S^{z}(\tilde {z})\sin\theta_i,\quad S_{i}^{y}= S_{i}^{y}(\tilde z)\quad S_{i}^{z}=-S^{x}(\tilde {z})\sin\theta_i+S^{z}(\tilde {z})\cos\theta_i,
 \end{align}
where $i$ is the sublattice index $i=a,b,c$, and $\theta_i=0,2\pi/3,4\pi/3$. The linear spin wave Hamiltonian can be expressed as~\cite{}
\begin{align}
\mathcal{H}^{(2)}=\frac{S}{2}\sum_{k}\Psi_{{k}}^{\dagger}H_{{k}}\Psi_{{k}},
\end{align}
where $\Psi_{k}=\{a_{k},a^{\dg}_{-k}\}^{T}$,
\begin{align}
H_{k}&=
\begin{pmatrix}
A_{k} & B_{k}\\
B_{k} & A_{-k}
\end{pmatrix},
\end{align}
$A_k=J(3+\frac{1}{2}\gamma_k)$, $B_k=-\frac{3}{2}J \gamma_k$, $\gamma_k=\cos k_x+2\cos k_x/2\cos\sqrt{3}k_y/2$. The spin wave Hamiltonian can be diagonalized by the transformation $\Psi_k=T\Psi'_{k}$, where $\Psi'_k=\{e_k,e^{\dg}_{-k}\}^T$, such that $T^{-1}\sigma_z H_k T=\omega_k\sigma_z$. We found $\omega_k=\sqrt{A_k^2-B_k^2}$, and
\begin{align}
T_k&=
\begin{pmatrix}
\sqrt{\frac{A_k+\omega_k}{2\omega_k}} & -\sgn{B_k}\sqrt{\frac{A_k-\omega_k}{2\omega_k}}\\
-\sgn{B_k}\sqrt{\frac{A_k-\omega_k}{2\omega_k}} & \sqrt{\frac{A_k+\omega_k}{2\omega_k}}
\end{pmatrix}.
\end{align}
As the $SU(2)$ spin rotation symmetry of the Hamiltonian is fully broken by the magnetic order, there are three Goldstone modes, associated with the three broken symmetry generators. It is straight forward to show that there are three zero modes at momentum $\Gamma=(0,0)$ and $\pm K= \pm(\frac{4\pi}{3},0)$, respectively. To check if they are the Goldstone modes and learn the spin structure, let us analyze the eigenmodes near $\Gamma,\pm K$. %In the following, we take $S=1/2$.

\textit{Near $\Gamma$} -- The spin wave spectrum is $\omega_{q+\Gamma}=\frac{3\sqrt{3}JS}{4}q$. The transformation matrix $T_k$ is
\begin{align}
T_{q+\Gamma}=
\frac{\sqrt[4]{3}}{ \sqrt{2q}}\times
\begin{pmatrix}
1&1\\
1 & 1
\end{pmatrix}+\frac{\sqrt{q}}{2\sqrt{2}\sqrt[4]{3}}\times
\begin{pmatrix}
1& -1\\
-1 & 1
\end{pmatrix}+\mathcal{O}(q)
\label{appeq:TGamma}
\end{align}
The part singular in $\sqrt{q}$ at order $1/\sqrt{q}$ corresponds to the Goldstone mode excitation, which contributes to the divergent static susceptibility as $q\rightarrow 0$, while the subleading term at order $\sqrt{q}$ corresponds to the soft modes, whose static susceptibility is finite at $q=0$. From the singular part, we obtain the leading order dynamical spin susceptibility (labeled by superscript ``(0)") at $S=1/2$
\begin{align}
-i \langle T S^{\tilde x}_{-q}(t) S^{\tilde x}_q(0)\rangle^{(0)}_{\omega}&=-\frac{i}{4} \langle T (a_{-q}(t)+a^{\dg}_{q}(t)) (a_q(0)+a^{\dg}_{-q}(0))\rangle_{\omega}=-i \langle T (e_{-q}(t)+e^{\dg}_{q}(t)) (e_q(0)+e^{\dg}_{-q}(0))\rangle_{\omega}\times \big(\frac{\sqrt[4]{3}}{ \sqrt{2q}}\big)^2\non\\
&=\frac{9JS}{2}\frac{1}{\omega^2-\omega_q^2}=\frac{9J}{4}\frac{1}{\omega^2-\omega_q^2}\non\\
-i \langle T S^{\tilde y}_{-q}(t) S^{\tilde y}_q(0)\rangle^{(0)}_{\omega}&=0.
\label{eq:GoldstoneGamma}
\end{align}
Similarly, we obtain from the $\sqrt{q}$ order terms in Eq.~\eqref{appeq:TGamma} the next order dynamical spin susceptibility (labeled by superscript ``(1)") at $S=1/2$
 \begin{align}
-i \langle T S^{\tilde x}_{-q}(t) S^{\tilde x}_q(0)\rangle^{(1)}_{\omega}=0,\quad\quad
-i \langle T S^{\tilde y}_{-q}(t) S^{\tilde y}_q(0)\rangle^{(1)}_{\omega}=\frac{3J q^2}{16}\frac{1}{\omega^2-\omega_q^2}.
\end{align}

Note that by $\tilde x, \tilde y$, we mean the spin component in the local coordinates. For our interest of obtaining the logarithmical divergent contribution to the thermal self-energy, only non-zero terms in Eq.~\eqref{eq:GoldstoneGamma} is needed, which physically means spin fluctuations along the local $\tilde x$ direction at the $\Gamma$ point (see Fig.~\ref{fig:SpinWaveCo}(b)).

\textit{Near $\pm K$} -- By doing the same analysis near $\pm K$, we found the spin wave spectrum is $\omega_{q\pm K}=\frac{3\sqrt{3}JS}{2\sqrt{2}}q$, and to the leading order in $q$,
\begin{align}
T_{q\pm K}
=\frac{\sqrt[4]{3}}{2^{3/4}\sqrt{q}}\times
\begin{pmatrix}
1& -1\\
-1 & 1
\end{pmatrix}+\mathcal{O}(\sqrt{q}).
\label{appeq:TK}
\end{align}
The leading order dynamical spin susceptibility at $S=1/2$ is
\begin{align}
-i \langle T S^{\tilde y}_{-q\pm K}(t) S^{\tilde y}_{q\mp K}(0)\rangle^{(0)}_{\omega}=\frac{9J}{8}\frac{1}{\omega^2-\omega_{q\pm K}^2}.
\label{appeq:GoldstoneK}
\end{align}
Eqs.~\eqref{eq:GoldstoneGamma} and \eqref{appeq:GoldstoneK} are essentially what we have in Eq.~\eqref{eq:Magnon} in the main text. The two Goldstone modes are shown graphically in Fig.~\ref{fig:SpinWaveCo}(c),(d).
%========================================================
\section{Calculation of $\mathcal{C}_n(z)$}\label{app:Fn}
To obtain $\mathcal{C}_n(z)$ for a generic $n$, we review the rules found in Sec.~\ref{sec:OneLoop}. These are
\begin{itemize}
\item We use $\circ$ and $\bullet$ for magnon-fermion vertex $\hat e\, \gamma^{v\dg}\gamma^c$ and  $\hat e\, \gamma^{c\dg}\gamma^v$, respectively. As vertices like $\hat e\, \gamma^{v\dg}\gamma^v$ are not considered to the leading logarithmical order, the renormalized Green's function at $n$-loop order should have $n$ pairs of alternating $\circ$ and $\bullet$ vertices;
\item Adding contributions from $\chi^{xx}$ and $\chi^{yy}$, each $\circ\aquarius\circ$ or $\bullet\aquarius\bullet$ magnon propagator contribute a factor $(\beta_1-2\beta_2)U^2$;
\item Similarly, each $\circ\aquarius\bullet$ or $\bullet\aquarius\circ$ magnon propagator contribute a factor $(\beta_1+2\beta_2)U^2$.
\end{itemize}
All diagrammatic configurations at $n$-loop order can be grouped by the total number of $\bullet\aquarius\bullet$, $\circ\aquarius\circ$, $\circ\aquarius\bullet$ and $\bullet\aquarius\circ$ propagators, and the contribution from each diagram is the same within a given group. In the following, we discuss $n=2m$ even and $n=2m+1$ odd seperately.
\subsection{n=2m}
Each diagram is in a group (labeled by $l$) that has $m-l$ of $\bullet\aquarius\bullet$, $m-l$ of $\circ\aquarius\circ$ propagators, and a total $2l$ of $\bullet\aquarius\circ$ or $\circ\aquarius\bullet$ propagators, where $l=0,1,...,m$. For a given group labeled as $l$, the combinatoric factor contributing to the renormalized Green's function is
\begin{align}
&\Big\{\Big[\frac{C_{2m}^{2}C_{2m-2}^{2}..C_{2l+2}^{2}}{(m-l)!}\Big]^2[(\beta_1-2\beta_2)U^2]^{2(m-l)}\Big\}\Big\{(2l)![(\beta_1+2\beta_2)U^2]^{2l}\Big\}\non\\
=&(\beta U^2)^{2m}\Big\{\Big[\frac{(2m)!}{(2m-2l)!!}\Big]^2\frac{z^{2l}}{(2l)!}\Big\},
\end{align}
where the first $\{...\}$ in the first line above comes from contributions of $\bullet\aquarius\bullet$ and $\circ\aquarius\circ$ propagators, and the second $\{...\}$ comes from contributions of $\bullet\aquarius\circ$ and $\circ\aquarius\bullet$ propagators. Summing up all factors, we find $\mathcal{C}_{2m}(z)=\sum_{l=0}^{m}\big [\frac{(2m)\,!}{(2m-2l)\,!!}\big]^2\frac{z^{2l}}{(2l)\,!}$.

\subsection{n=2m+1}
For $n$ odd, as the total number of $\circ$ or $\bullet$ vertices used up for $\circ\aquarius\circ$ or $\bullet\aquarius\bullet$ propagators must be even, there must be a total odd number of $\bullet\aquarius\circ$ and $\circ\aquarius\bullet$ propagators. As a result, each group labeled by $l$ has $m-l$ of $\bullet\aquarius\bullet$, $m-l$ of $\circ\aquarius\circ$ propagators, and a total $2l+1$ of $\bullet\aquarius\circ$ and $\circ\aquarius\bullet$ propagators, where $l=0,1,...,m$. The combinatoric factor is
\begin{align}
&\Big\{\Big[\frac{C_{2m+1}^{2}C_{2m-1}^{2}..C_{2l+3}^{2}}{(m-l)!}\Big]^2[(\beta_1-2\beta_2)U^2]^{2(m-l)}\Big\}\Big\{(2l+1)![(\beta_1+2\beta_2)U^2]^{2l+1}\Big\}\non\\
=&(\beta U^2)^{2m+1}\Big\{\Big[\frac{(2m+1)!}{(2m-2l)!!}\Big]^2\frac{z^{2l+1}}{(2l+1)!}\Big\}.
\end{align}
Summing up all factors, we find $\mathcal{C}_{2m+1}(z)=\sum_{l=0}^{m}\big [\frac{(2m+1)\,!}{(2m-2l)\,!!}\big]^2\frac{z^{2l+1}}{(2l+1)\,!}$.

%========================================================
\section{Evaluate the spectral function}\label{app:SpectralFunction}
\begin{figure}[bth!]
\centering
\subfigure[]{\includegraphics[width=0.4\columnwidth]{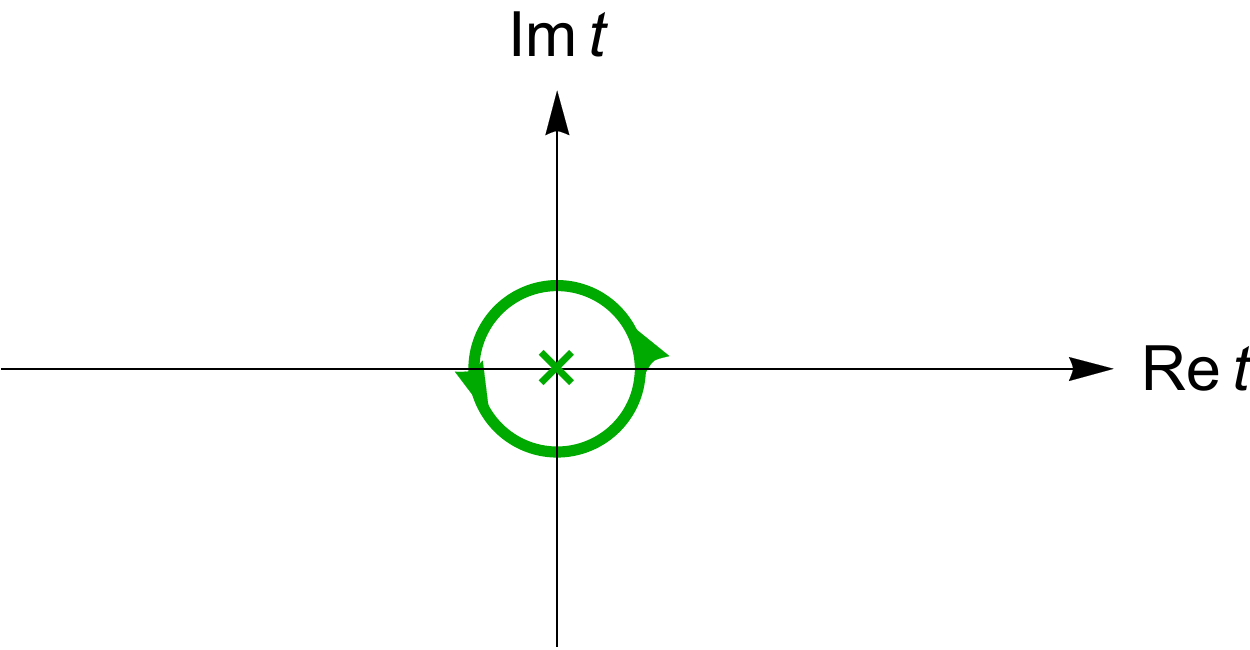}}\quad\quad
\subfigure[]{\includegraphics[width=0.4\columnwidth]{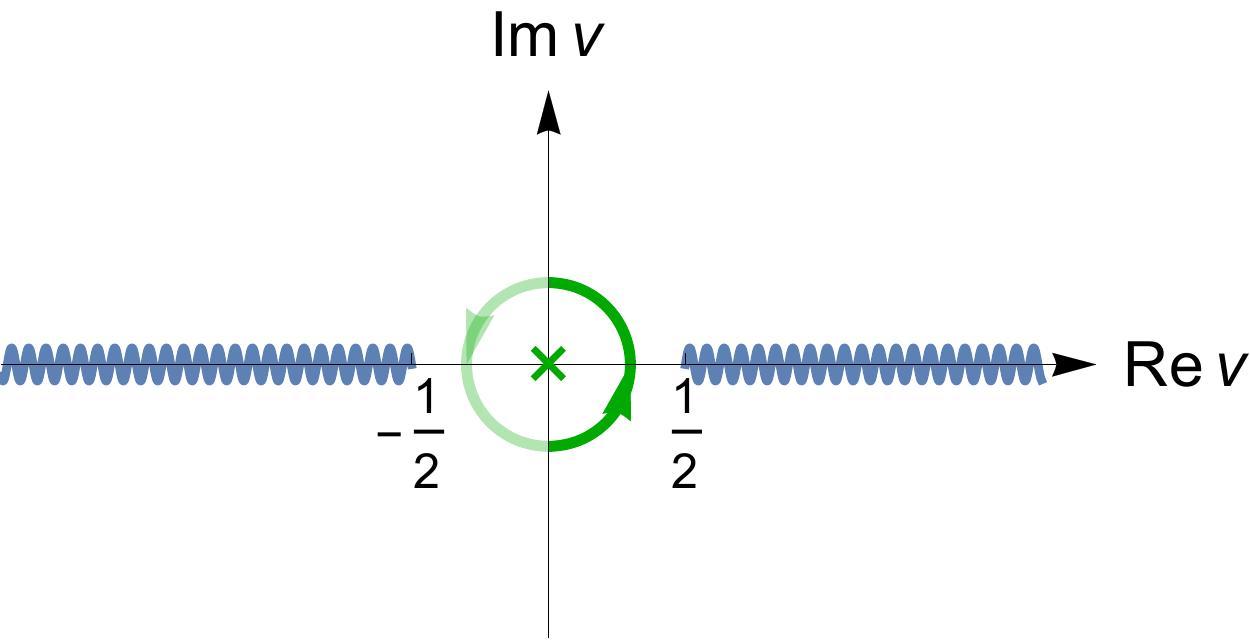}}
\subfigure[]{\includegraphics[width=0.4\columnwidth]{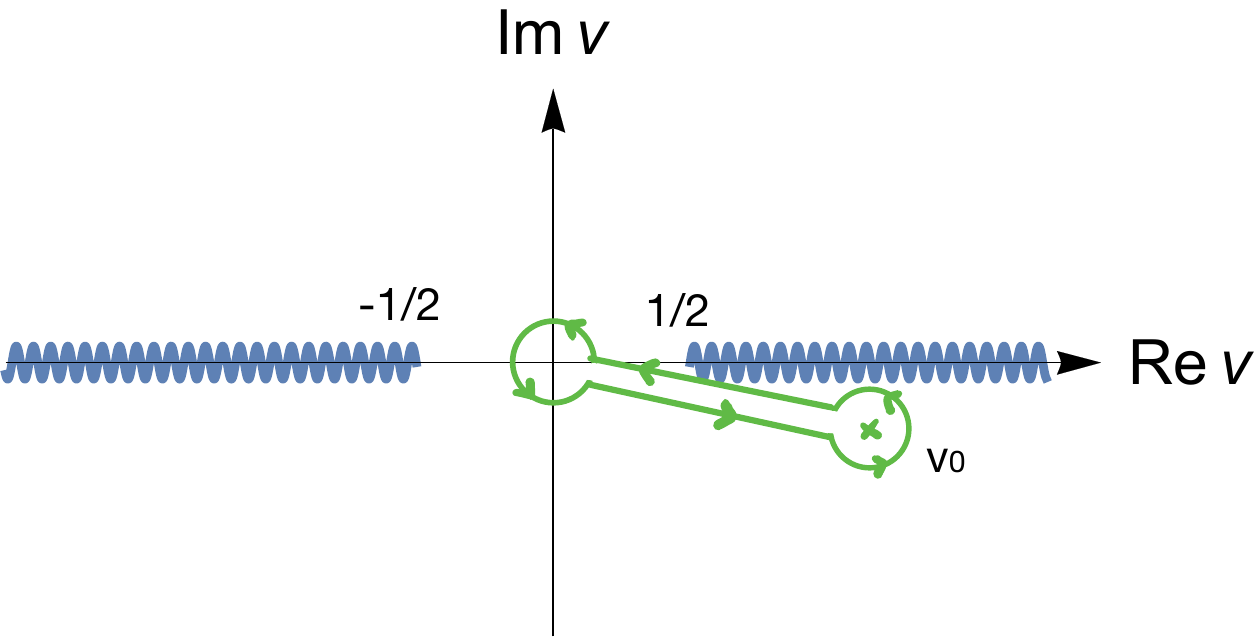}}\quad\quad\quad
\subfigure[]{\includegraphics[width=0.38\columnwidth]{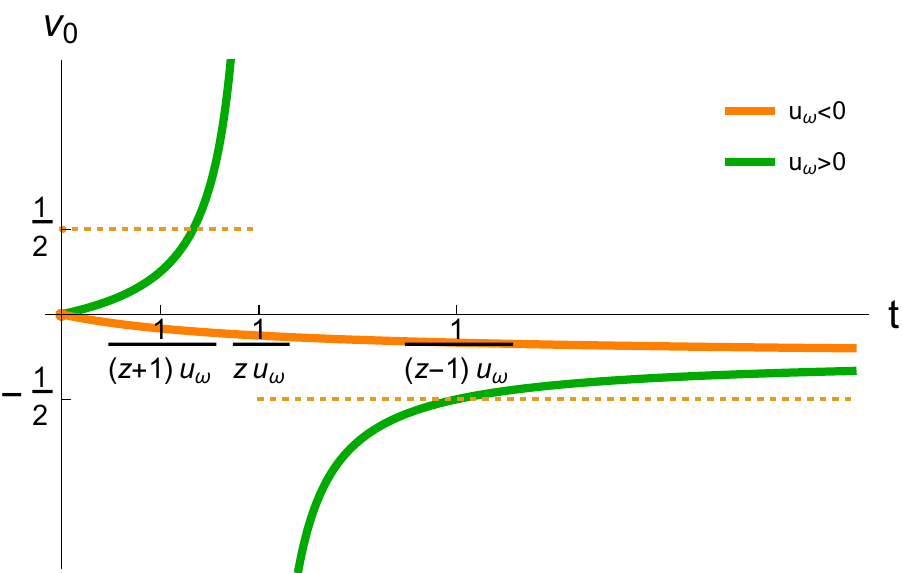}}
\caption{(a)-(c): The integration contours for the computation of the combinatoric factors. (a) The integration contour for Eqs.~\eqref{eq:contour1} and \eqref{eq:Fexpression3}. There is only one multi-pole at $t=0$ for each given $n$ and $l$. (b) 
The contour for Eqs.~\eqref{eq:Fexpression4} and \eqref{eq:Fexpression5}. The contour contains the multi-pole and the branch cuts (blue wavy lines). The parts of the contour on the right (darker green line) and on the left (lighter green line)  come from the first and second terms in Eq.~\eqref{eq:sum2D}. (c) The integration contour for Eq.~\eqref{eq:sum1}. The multi-pole at $v=0$ moves and becomes a single pole at $v_0$. (d) $v_0$ as a function of $t\in (0,\infty)$ for $u_{\omega}>0$ and $u_{\omega}<0$.} \label{fig-11}
\end{figure}

We now evaluate the spectral function defined as $A^{c,v}(\khs,\omega)=-\frac{1}{\pi}\im G^{c,v}(k_{hs},\omega+i\delta)$ analytically starting from Eqs.~\eqref{eq:Fexpression} and \eqref{eq:FullGreen}.
%, and discuss its qualitative features as $z$ and the order parameter $\Delta$ vary.
The key challenge is to perform the summation over $n$ in Eq.~\eqref{eq:FullGreen} where $\mathcal{C}_n(z)$ doesn't have a simple closed form. Moreover, as $\mathcal{C}_{n}(z)\sim \mathcal{O}(n!)$, a numerical calculation of $A^{c,v}(\khs,\omega)$ is quite challenging on its own. Our point of departure is to sum over $l$ in Eq.~\eqref{eq:Fexpression} by noting that from $\frac{(p+q)\,!}{p\,! q\,!}=\frac{1}{2\pi i}\oint^{(0+)}\diff t\, t^{-p-1}(1-t)^{-q-1}$ [$p,q\in \text{Integers}\, (\mathbb Z)$ and $p+q\geq -2$]~\cite{SpecialFunction},
\begin{align}\label{eq:contour1}
\frac{1}{[(2m-2l)\,!!]^2}=\frac{1}{2^{2m-2l}}\frac{1}{(2m-2l)\,!}\frac{1}{2\pi i}\oint^{(0+)}\diff t\, t^{-m+l-1}(1-t)^{-m+l-1},
\end{align}
where $\oint^{(0+)}$ means the contour integral goes around the pole at $t=0$ counter-clockwisely [see Fig.~\ref{fig-11}(a)]. For concreteness, we take $n=2m$ as an example. Plug Eq.~\eqref{eq:contour1} into Eq.~\eqref{eq:Fexpression}, we have
\begin{align}\label{eq:Fexpression3}
\mathcal{C}_{2m}(z)=\frac{(2m)\,!}{2^{2m}}\frac{1}{2\pi i}\oint^{(0+)}\diff t\, \big[\frac{1}{t(1-t)}\big]^{m+1}\sum_{l=0}^{m}\frac{(2m)\,!}{(2m-2l)\,!(2l)\,!}(2z)^{2l}[t(1-t)]^{l}.
\end{align}
To sum over $l$ in Eq.~\eqref{eq:Fexpression3}, we note that from $(1+x)^m=\sum_{p=0}^{m}\frac{m!}{(m-p)!\, p!}x^p$,
\begin{align}\label{eq:sum2D}
&\sum_{l=0}^{m}\frac{(2m)\,!}{(2m-2l)\,!(2l)\,!}(2z)^{2l}[t(1-t)]^{l}=\sum_{l=0}^{m}\frac{(2m)\,!}{(2m-2l)\,!(2l)\,!}(2z)^{2l}v^{2l}\non\\
=& \frac{1}{2}\{\sum_{p=0}^{2m}\frac{(2m)!}{(2m-p)!\,p!}(2zv)^{p}+\sum_{p=0}^{2m}\frac{(2m)!}{(2m-p)!\,p!}(-2zv)^{p}\}\non\\
=&\frac{1}{2}\big[(1+2zv)^{2m}+(1-2zv)^{2m}\big],
\end{align}
where we define $v=\sqrt{t(1-t)}$. Note that by summing over $l$, non-analytic branch-cuts must be introduced, which turns important to get the imaginary part of the spectral function later. Changing variable from $t$ to $v$, the integration contour changes from a circle around $t=0$ to a semi-circle around $v=0$ on the right-half-plane, and $\diff t=\frac{2v}{\sqrt{1-4v^2}}\diff v$. By adding the two terms in Eq.~\eqref{eq:sum2D} and changing variable $v\rightarrow -v$ for the second term, Eq.~\eqref{eq:Fexpression3} becomes [see Fig.~\ref{fig-11}(b)]
\begin{align}\label{eq:Fexpression4}
\mathcal{C}_{2m}(z)=\frac{(2m)\,!}{2^{2m}}\frac{1}{2\pi i}\oint^{(0+)}\diff v\, \big(\frac{1}{v}\big)^{2m+2}\frac{v}{\sqrt{1-4v^2}}(1+2zv)^{2m}.
\end{align}
Similarly, we found that $\mathcal{C}_{2m+1}(z)$ has the same form but changes $2m$ in the expression to $2m+1$, thus
\begin{align}\label{eq:Fexpression5}
\mathcal{C}_{n}(z)=\frac{n!}{2^{n}}\frac{1}{2\pi i}\oint^{(0+)}\diff v\, \big(\frac{1}{v}\big)^{n+2}\frac{v}{\sqrt{1-4v^2}}(1+2zv)^{n} \text{ for } n\in \mathbb Z.
\end{align}
From Eq.~\eqref{eq:Fexpression3} to Eq.~\eqref{eq:Fexpression5}, we are essentially transforming the summation over $l$ in Eq.~\eqref{eq:Fexpression3} into evaluating the residue of the integrand at $v=0$ in Eq.~\eqref{eq:Fexpression5}. The gain we have is that the number $n$ now only appears as a simple coefficient $n!$ and as exponents in the integrand, which simplifies the summation over $n$ in Eq.~\eqref{eq:FullGreen}.

To find $A^{c,v}(\khs,\omega)$, we analytically continue $G^{c,v}(\khs,\omega)$ from Eq.~\eqref{eq:FullGreen} to real frequencies by replacing $i\omega\rightarrow \omega+i\delta$. As long as the series over $n$ converge, doing the analytical continuation before or after the summation over $n$ should give the same result. For convenience, we perform the analytical continuation before the summation, and have
\begin{align}\label{eq:spectral}
A^{c,v}(\khs,\omega)&=-\frac{1}{\pi}\im G^{c,v}(k_{hs},\omega+i\delta)\non\\
&=-\frac{1}{\pi}\im \{G^{c,v\,(0)}(\khs,\omega+i\delta)\sum_{n=0}^{\infty}\mathcal{C}_n(z)[u(\omega+i\delta)]^{n}\},
\end{align}
where $u(\omega+i\delta)=\beta U^2 G^{v,c\,(0)}(\khs,\omega+i\delta)G^{c,v\,(0)}(\khs,\omega+i\delta)$. The imaginary part inside $\{...\}$ comes from two places --  from $\im G^{c,v\,(0)}(\khs,\omega+i\delta)\times\re \sum_{n=0}^{\infty} ...$ and from $\re G^{c,v\,(0)}(\khs,\omega+i\delta)\times\im \sum_{n=0}^{\infty} ...$. As $\im G^{c,v\,(0)}(\khs,\omega+i\delta)=-i\pi\delta(\omega\mp \Delta)$, the first contribution should be a delta-function peak at $\omega=\pm \Delta$ if $\int_{\Delta-0}^{\Delta+0}\diff \omega A(\omega)$ is finite. We checked that this integral actually vanishes. This implies that thermal fluctuations destroy the delta-function peak. To evaluate the second contribution, we use Eq.~\eqref{eq:Fexpression5} and $n!=\int_{0}^{+\infty}\diff t \,e^{-t}t^n$, express $\sum_{n=0}^{\infty}\mathcal{C}_n(z)[u(\omega+i\delta)]^{n}$ in Eq.~\eqref{eq:spectral} as
\begin{align}\label{eq:sum1}
\sum_{n=0}^{\infty}\mathcal{C}_n(z)u_{\omega}^{n}&=\frac{1}{2\pi i}\sum_{n=0}^{\infty}n!\,\oint^{(0+)}\diff v\, \frac{1}{v}\frac{1}{\sqrt{1-4v^2}}\Big[\frac{u_{\omega}(1+2zv)}{2v}\Big]^{n}\non\\
&=\frac{1}{2\pi i}\int_{0}^{+\infty}\diff t\,e^{-t}\oint^{(0+)}\diff v\, \frac{1}{v}\frac{1}{\sqrt{1-4v^2}}\sum_{n=0}^{\infty}\Big[\frac{u_{\omega}\,t\,(1+2zv)}{2v}\Big]^{n}\non\\
&=\frac{1}{2\pi i}\int_{0}^{+\infty}\diff t\,e^{-t}\oint\diff v\, \frac{1}{v}\frac{1}{\sqrt{1-4v^2}}\frac{1}{1-\frac{u_{\omega}\,t\,(1+2zv)}{2v}}\non\\
&=\frac{1}{2\pi i}\int_{0}^{+\infty}\diff t\,e^{-t}\oint^{(0-v_0,+)}\diff v\, \frac{2}{\sqrt{1-4v^2}}\frac{1}{2v-u_{\omega}\,t\,(1+2zv)}\non\\
&=\frac{1}{2\pi i}\int_{0}^{+\infty}\diff t\,e^{-t}\oint^{(0-v_0,+)}\diff v\, \frac{2}{\sqrt{1-4v^2}}\frac{1}{2(1-u_{\omega}\,t\,z)(v-\frac{u_{\omega}\,t}{2(1-u_{\omega}\, t\, z)})},
\end{align}
where $u(\omega+i\delta)$ is replaced by $u_{\omega}$ for brevity. Importantly, by summing over $n$, the multi-pole at $v=0$ vanishes, while a single pole at $v=v_0=\frac{u_{\omega}\,t}{2(1-u_{\omega}\, t\, z)}$ emerges due to the non-analyticity at $v=0$. Then the contour $\oint^{(0+)}$ changes to $\oint^{(0-v_0,+)}$, where $\oint^{(0-v_0,+)}$ indicates a counter-clockwise contour enclosing $v=0$ and $v=v_0$ [see Fig.~\ref{fig-11}(c)]. Enforcing $\omega\rightarrow \omega+i\delta$, we find $u_{\omega}\rightarrow u_{\omega}-i\delta \sgn(\omega)$ and $v_0\rightarrow v_0-i\delta \sgn(\omega)$. As $v_0$ remains on the lower or upper half-plane [depending on $\sgn{(\omega)}$] as $t$ varies, the integrals from $0$ to $v_0$ and from $v_0$ to $0$ cancel, and Eq.~\eqref{eq:sum1} becomes
\begin{align}\label{eq:sum2}
\sum_{n=0}^{\infty}\mathcal{C}_n(z)u_{\omega}^{n}=\frac{1}{2\pi i}\int_{0}^{+\infty}\diff t\,e^{-t}\oint^{(v_0+)}\diff v\, \frac{1}{\sqrt{1-4v^2}}\frac{1}{(1-u_{\omega}\,t\,z)(v-v_0)}.
\end{align}
To obtain the imaginary part of Eq.~\eqref{eq:sum2}, one can show that the only contribution comes from the residue of the integrand of $\oint^{(v_0+)}$ when $v_0$ sits at the branch cut, i.e., $|v_0|\geq 1/2$. By examining $v_0=\frac{u_{\omega}\,t}{2(1-u_{\omega}\, t\, z)}$ for $t\in (0,\infty)$, we find $|v_0|\geq 1/2$ only when $u_{\omega}>0$ and $t\in (\frac{1}{(z+1)u_{\omega}},\frac{1}{(z-1)u_{\omega}})$ [see Fig.~\ref{fig-11}(d)]. In particular, if $z=1$, the upper bound for $t$ is $+\infty$.
Thus the imaginary part of Eq.~\eqref{eq:sum2} when $u_{\omega}>0$ is
\begin{align}\label{eq:sum3}
i\im \sum_{n=0}^{\infty}\mathcal{C}_n(z)u_{\omega}^{n}&=\int_{\frac{1}{(z+1)u_{\omega}}}^{\frac{1}{(z-1)u_{\omega}}}\diff t\, e^{-t}\frac{1}{2\pi i}\oint^{(v_0+)}\diff v\, \frac{1}{\sqrt{1-4v^2}}\frac{1}{(1-u_{\omega}\,t\,z)(v-v_0)}\non\\
&=\int_{\frac{1}{(z+1)u_{\omega}}}^{\frac{1}{(z-1)u_{\omega}}}\diff t\, e^{-t}\, \frac{1}{\sqrt{1-4v_0^2}}\frac{1}{1-u_{\omega}\,t\,z}
\end{align}

In the following, let us consider $\omega>0$ for concreteness, so $v_0\rightarrow v_0-i\delta$. Note that $\frac{1}{\sqrt{1-4v_0^2}}$ is pure imaginary when $t\in (\frac{1}{(z+1)u_{\omega}},\frac{1}{(z-1)u_{\omega}})$. As we explain below, it needs some care to determine the sign at different $t$. From Fig.~\ref{fig-11}(d), we see that when $t\in (\frac{1}{(z+1)u_{\omega}},\frac{1}{ z u_{\omega}})$, $\re v_0>1/2$, so $\im \frac{1}{\sqrt{1-4 v_0^2}}=\frac{1}{\sqrt{1+2 v_0}\sqrt{1-2v_0}}=\frac{-i}{\sqrt{1+2v_0}\sqrt{2v_0-1}}=\frac{-i}{\sqrt{4v_0^2-1}}$, $1-u_{\omega}\,t\,z>0$; when $t\in (\frac{1}{z\, u_{\omega}},\frac{1}{ (z-1) u_{\omega}})$, $\re v_0<-1/2$, so $\im \frac{1}{\sqrt{1-4 v_0^2}}=\frac{1}{\sqrt{1+2 v_0}\sqrt{1-2v_0}}=\frac{i}{\sqrt{4v_0^2-1}}$, $1-u_{\omega}\,t\,z<0$. So Eq.~\eqref{eq:sum3} becomes
\begin{align}
i\im \sum_{n=0}^{\infty}\mathcal{C}_n(z)u_{\omega}^{n}&=\int_{\frac{1}{(z+1)u_{\omega}}}^{\frac{1}{z \,u_{\omega}}}\diff t\, e^{-t}\, \frac{-i}{\sqrt{4v_0^2-1}}\frac{1}{1-u_{\omega}\,t\,z}+\int_{\frac{1}{z\,u_{\omega}}}^{\frac{1}{(z-1) \,u_{\omega}}}\diff t\, e^{-t}\, \frac{i}{\sqrt{4v_0^2-1}}\frac{1}{1-u_{\omega}\,t\,z}\non\\
&=\int_{\frac{1}{(z+1)u_{\omega}}}^{\frac{1}{(z-1) \,u_{\omega}}}\diff t\, e^{-t}\, \frac{-i}{\sqrt{(4v_0^2-1)(1-u_{\omega}\,t\,z)^2}}\non\\
&=-i\int_{\frac{1}{(z+1)u_{\omega}}}^{\frac{1}{(z-1) \,u_{\omega}}}\diff t\, e^{-t}\, \frac{1}{\sqrt{(u_{\omega}\,t)^2-(1-u_{\omega}\,t\,z)^2}},
\end{align}
where the integral over $t$ is convergent and positive definite. Similarly, we find when $\omega<0$, $i\im \sum_{n=0}^{\infty}\mathcal{C}_n(z)u_{\omega}^{n}=i\int_{\frac{1}{(z+1)u_{\omega}}}^{\frac{1}{(z-1) \,u_{\omega}}}\diff t\, e^{-t}\, \frac{1}{\sqrt{(u_{\omega}\,t)^2-(1-u_{\omega}\,t\,z)^2}}$. Plug them back to Eq.~\eqref{eq:spectral}, the spectral function is
\begin{align}\label{eq:result0}
A^{c,v}(\khs,\omega)=\frac{1}{\pi}\Big|\frac{1}{\omega\mp \Delta}\Big|\int_{\frac{1}{(z+1)u_{\omega}}}^{\frac{1}{(z-1) \,u_{\omega}}}\diff t\, e^{-t}\, \frac{1}{\sqrt{(u_{\omega}\,t)^2-(1-u_{\omega}\,t\,z)^2}}\Theta(u_{\omega}),
\end{align}
where we remind $u_{\omega}=\frac{\beta U^2}{\omega^2-\Delta^2}$.

\bibliography{Pseudogap_Thermal}
\end{document}